\documentclass[conference]{IEEEtran}
\IEEEoverridecommandlockouts
\usepackage{cite}
\usepackage{amsmath,amssymb,amsfonts}
\usepackage{algorithmic}
\usepackage{graphicx}
\usepackage{textcomp}
\usepackage{xcolor}
\usepackage{enumitem}
\usepackage{float}
\usepackage{adjustbox}
\usepackage{algorithm}
\usepackage{multirow}
\usepackage{tabularx}
\usepackage{url}
\usepackage{booktabs}
\usepackage{hyperref}
\usepackage{array}
\usepackage{framed}
\usepackage{caption}

\usepackage[table]{xcolor}

\definecolor{shadecolor}{rgb}{0.9,0.9,0.9}
\newcommand{\vpara}[1]{\vspace{0.05in}\noindent \textbf{#1 }}
\newcommand{\model}{\textsc{LLMIA}}
\newcommand{\smodel}{\textsc{LLMIA} }
\newcommand{\modelabb}{\textsc{LLMIA}}

\def\BibTeX{{\rm B\kern-.05em{\sc i\kern-.025em b}\kern-.08em
    T\kern-.1667em\lower.7ex\hbox{E}\kern-.125emX}}

\begin{document}

\title{\model: An Out-of-the-Box Index Advisor via In-Context Learning with LLMs}

\author{
    \IEEEauthorblockN{
        Xinxin Zhao\IEEEauthorrefmark{1},
        Xinmei Huang\IEEEauthorrefmark{1},
        Haoyang Li\IEEEauthorrefmark{1},
        Jing Zhang\IEEEauthorrefmark{1}\IEEEauthorrefmark{2},\\
        Shuai Wang\IEEEauthorrefmark{3},
        Tieying~Zhang\IEEEauthorrefmark{3},
        Jianjun Chen\IEEEauthorrefmark{3},
        Rui Shi\IEEEauthorrefmark{3},
        Cuiping Li\IEEEauthorrefmark{1},
        Hong Chen\IEEEauthorrefmark{1}
    }
    \IEEEauthorblockA{
        \IEEEauthorrefmark{1}School of Information, Renmin University of China, Beijing, China,
        \IEEEauthorrefmark{3}ByteDance Inc., China
    }
    \IEEEauthorblockA{
        \IEEEauthorrefmark{1}\{zhaoxinxin798, huangxinmei, lihaoyang.cs, zhang-jing, licuiping, chong\}@ruc.edu.cn,\\
        \IEEEauthorrefmark{3}\{wangshuai.will, tieying.zhang, jianjun.chen, shirui\}@bytedance.com
    }
    \thanks{\IEEEauthorrefmark{2} Jing Zhang is the corresponding author.}
}


\maketitle

\begin{abstract}
Index recommendation is crucial for optimizing database performance. However, existing heuristic- and learning-based methods often rely on inefficient exhaustive search and estimated costs, leading to low efficiency (due to the vast search space) and unsatisfactory actual latency (due to inaccurate estimations). Inspired by the refinement strategies of experienced DBAs—who efficiently identify and iteratively refine indexes with database feedback—we present \model, an out-of-the-box, tuning-free index advisor leveraging large language models (LLMs) through in-context learning for index recommendation. \smodel injects database expertise into the LLM using a high-quality demonstration pool and comprehensive workload feature extraction, while iteratively incorporating database feedback to guide the index refinement. This design enables \smodel to emulate the decision-making process of expert DBAs: efficiently recommending and refining indexes for various workloads within just a few interactions with the DBMS. 
We validate \smodel with extensive experiments on five standard OLAP benchmarks (TPC-H with different scales, JOB, TPC-DS, SSB), where it consistently outperforms or matches 12 baselines by producing superior index recommendations with minimal database interactions. Additionally, \smodel demonstrates robust generalization on two real-world commercial workloads, delivering high-quality recommendations without the need for additional adaptation or retraining, highlighting its out-of-the-box capability.

\end{abstract}


\newcommand{\figlc}{
\begin{figure}[t]
    \vspace{-1em}
    \centering
    \includegraphics[width=0.4\textwidth]{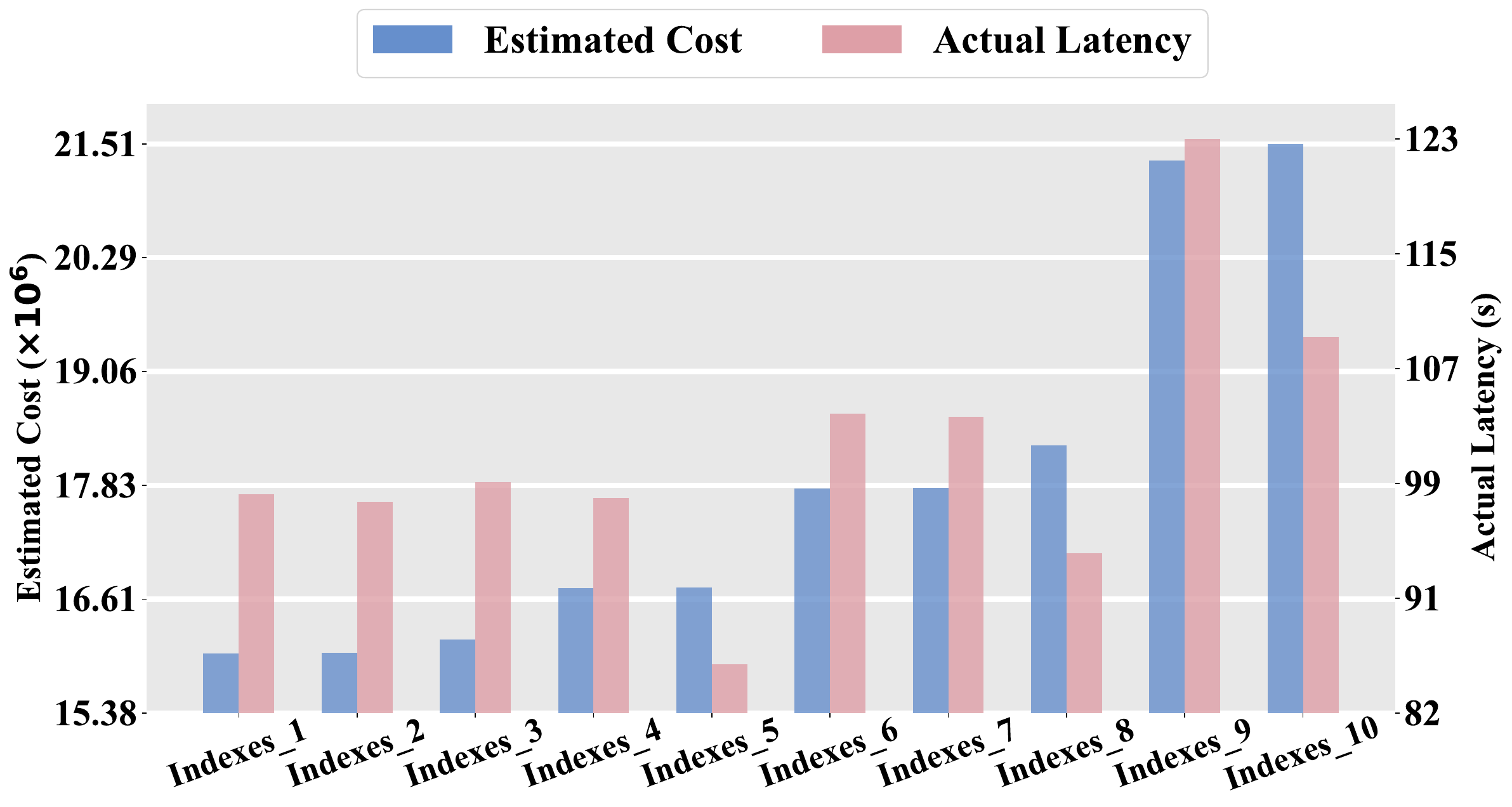}
	\caption{\label{fig:lc} Estimated cost vs. actual latency on TPC-H with different indexes. Estimated cost does not always correlate with actual latency, leading to sub-optimal index recommendations.}
\end{figure}
}

\newcommand{\figoverview}{
\begin{figure*}[t]
    \centering
    \includegraphics[width=0.70\textwidth]{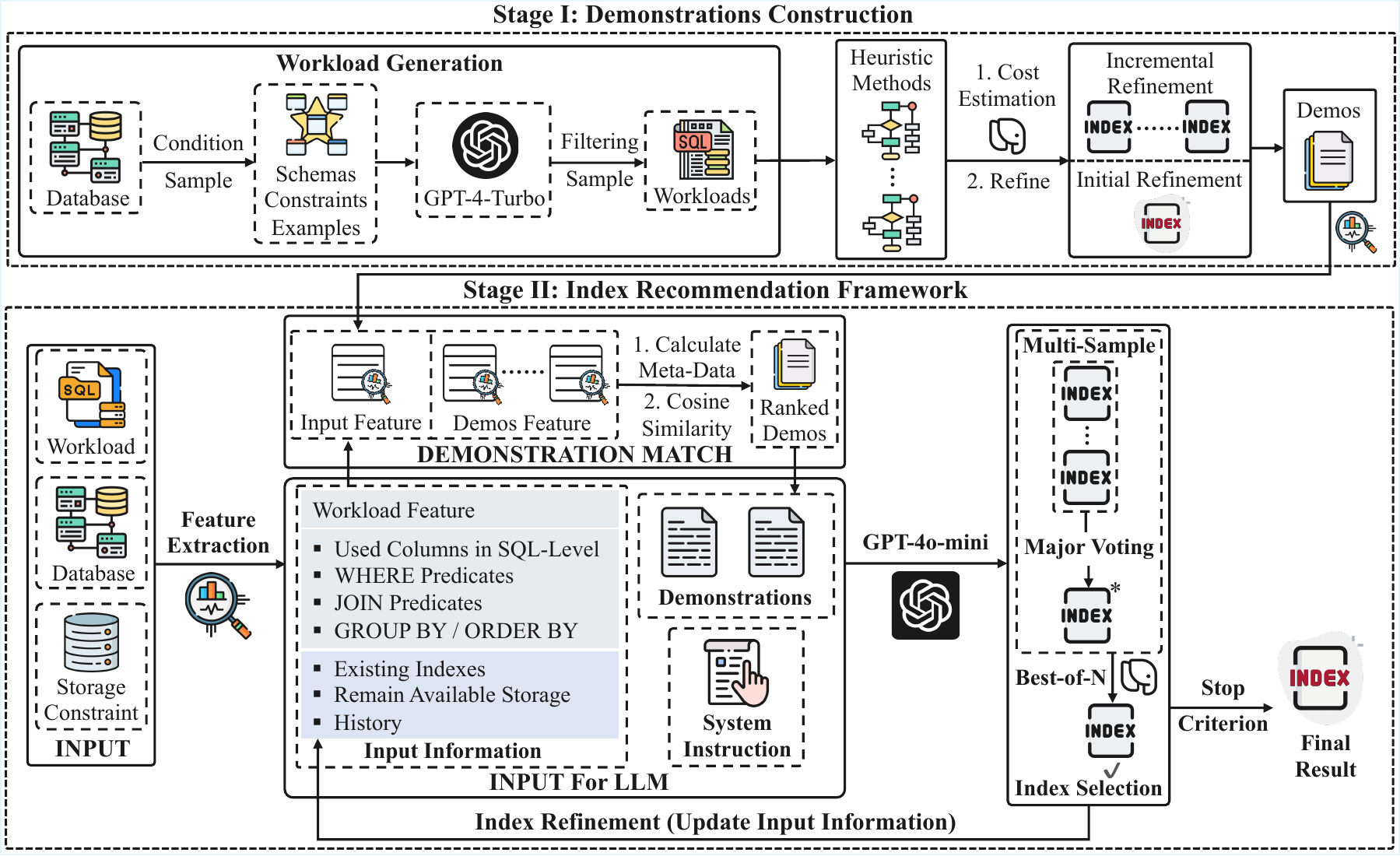}
    \caption{\label{fig:overview} Overview of \model, involving the Demonstration Construction and the Index Recommendation Framework.}
\end{figure*}
}

\newcommand{\figdc}{
\begin{figure}[t]
    \centering
    \includegraphics[width=0.42\textwidth]{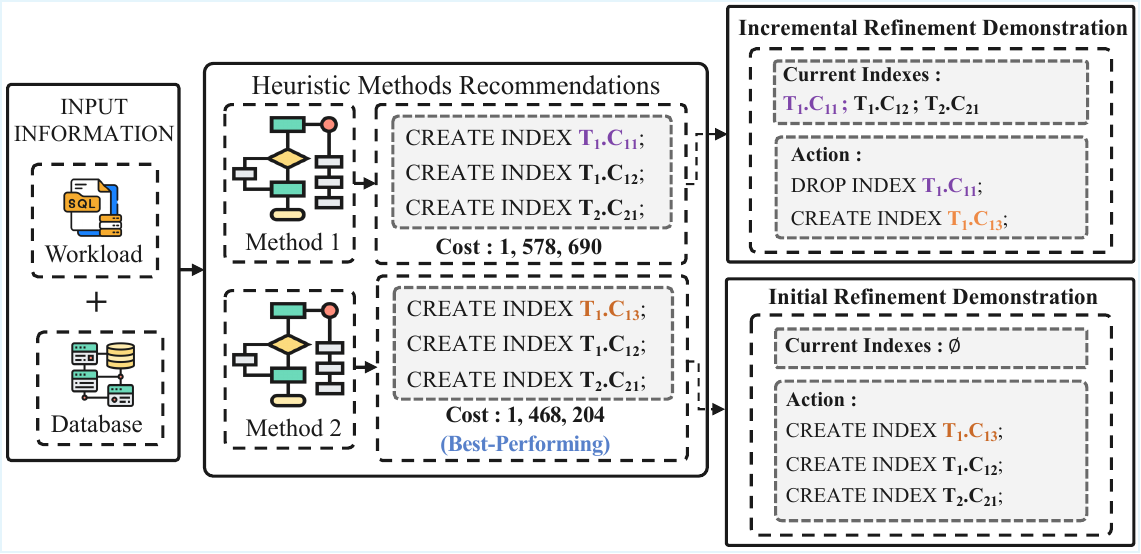}
	\caption{\label{fig:dc} Illustration of the refinement demonstration. For clarity, only two heuristic methods are shown as examples.}
\end{figure}
}

\newcommand{\figmv}{
\begin{figure}[]
    \centering
    \includegraphics[width=0.44\textwidth]{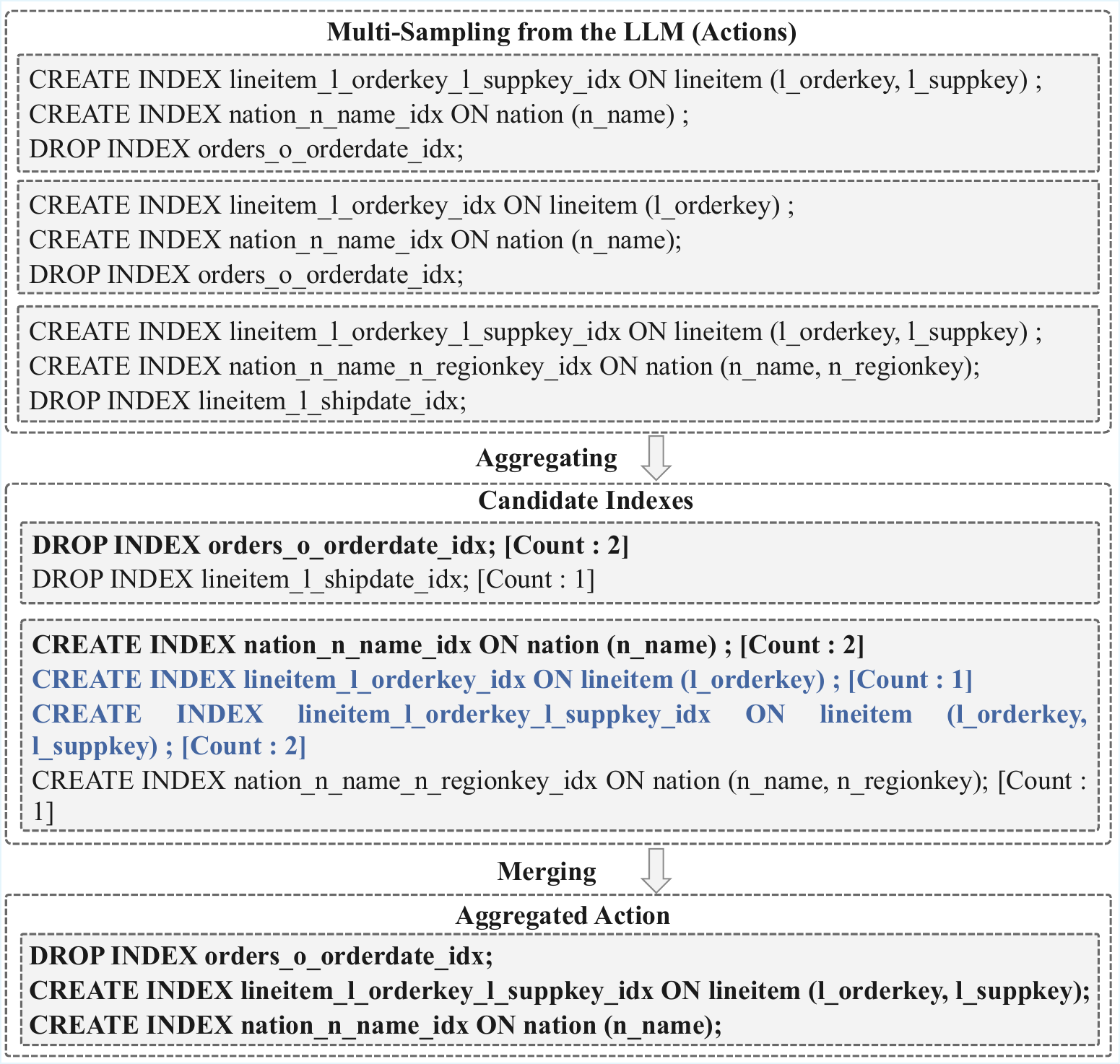}
	\caption{\label{fig:mv} Illustration of ``Index-Guided Major Voting''.}
\end{figure}
}

\newcommand{\figmainresults}{
\begin{figure*}[t]
    \centering
    \includegraphics[width=0.92\textwidth]{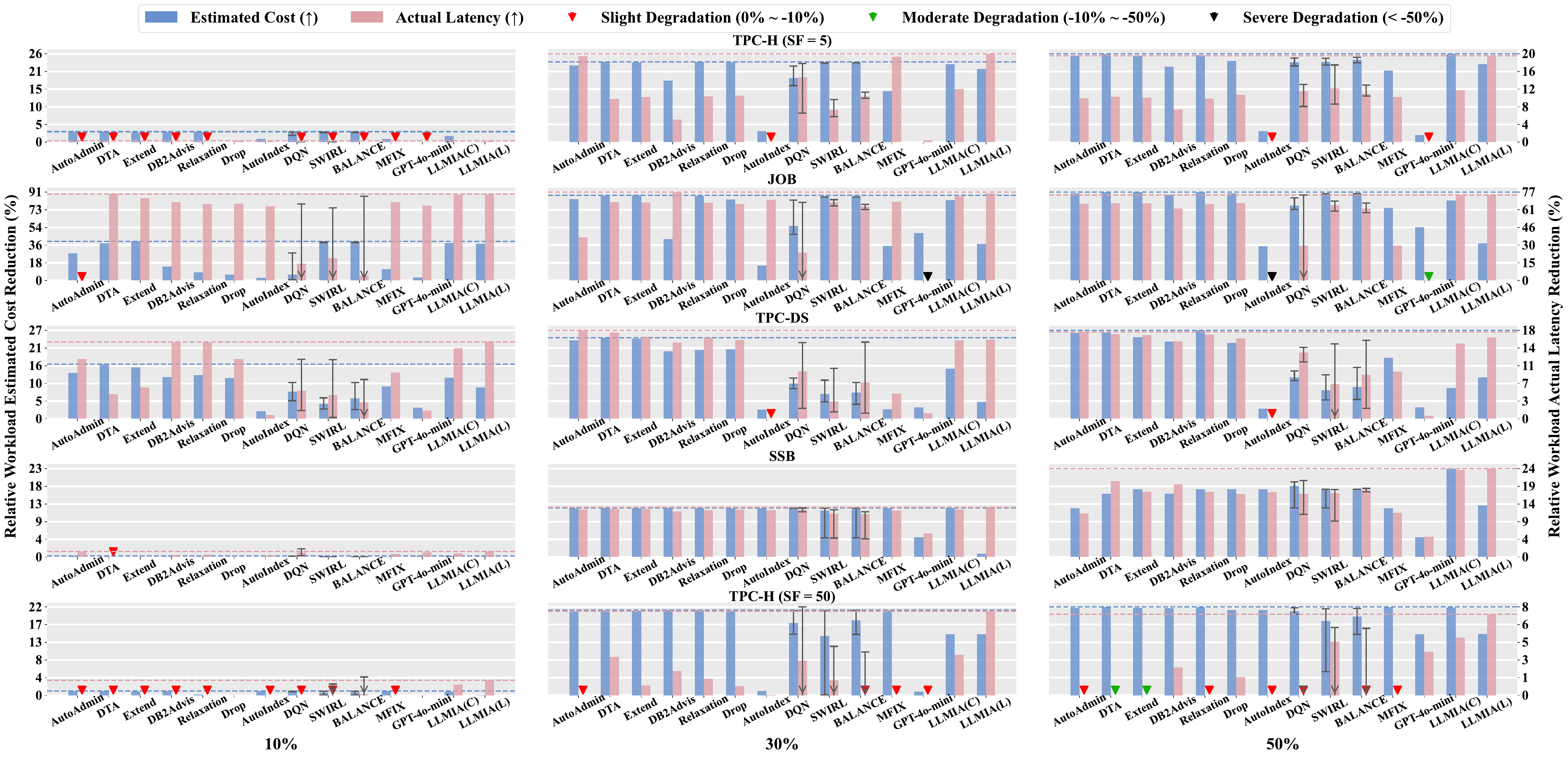}
	\caption{\label{fig:mainresults} Workload performance evaluation across 5 OLAP benchmarks under different storage constraints.}
\end{figure*}
}

\newcommand{\figrealworld}{
\begin{figure}[h]
    \centering
    \includegraphics[width=0.46\textwidth]{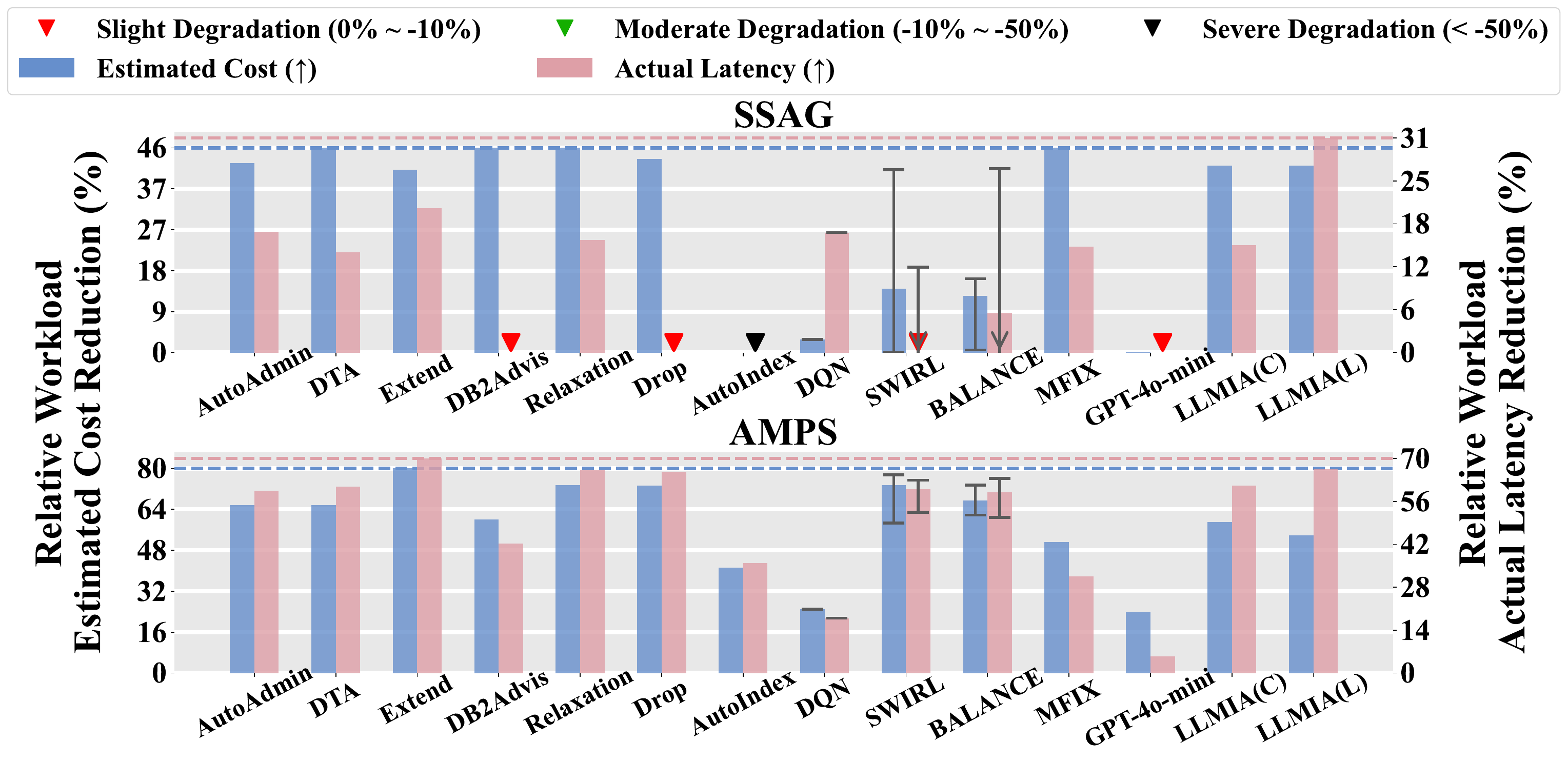}
	\caption{\label{fig:rw} Workload performance evaluation across 2 real-world benchmarks under the storage constraint of 30\%.} 
\end{figure}
}

\newcommand{\figlat}{
\begin{figure}[h]
    \centering
    \includegraphics[width=0.46\textwidth]{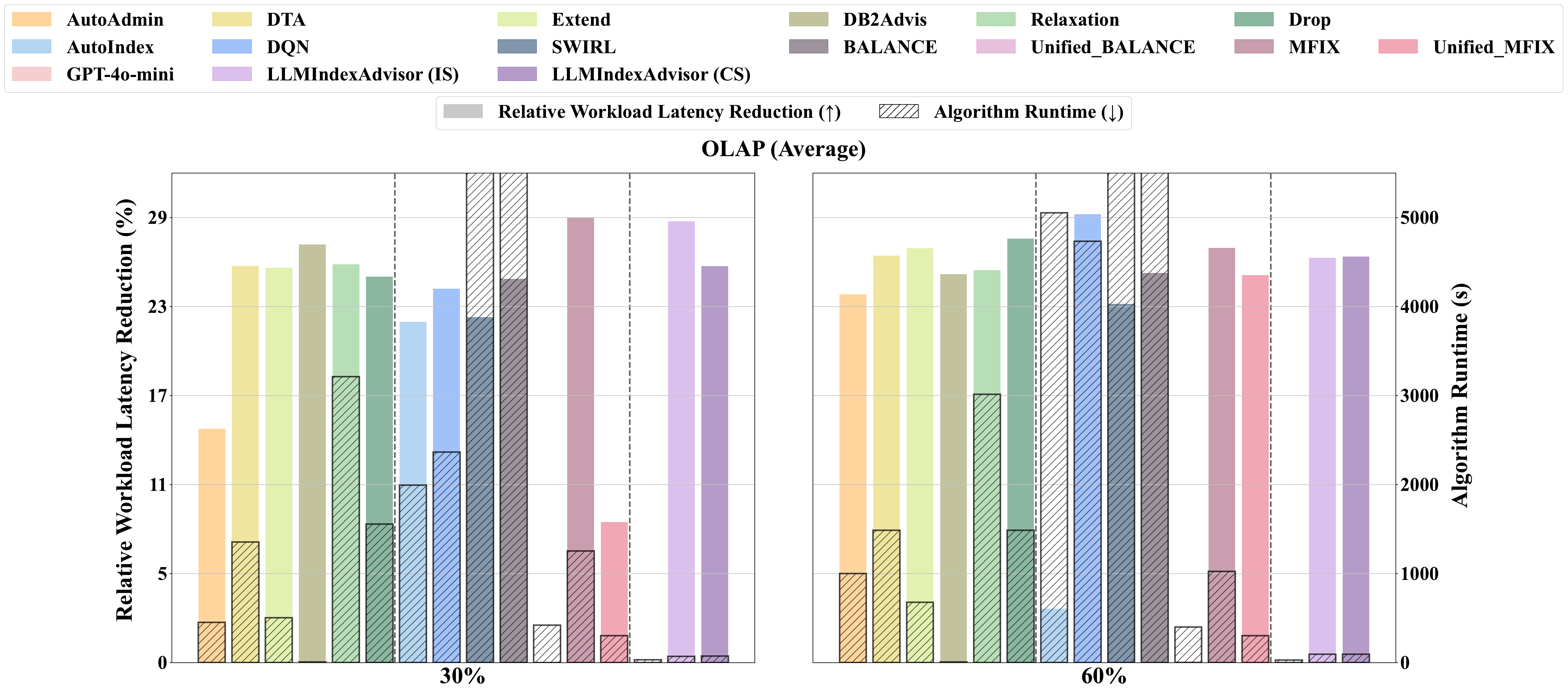}
	\caption{\label{fig:lat} Average workload latency evaluation.}
\end{figure}
}

\newcommand{\figsyspromptapp}{
\begin{figure*}[h]
    \centering
    \includegraphics[width=0.8\textwidth]{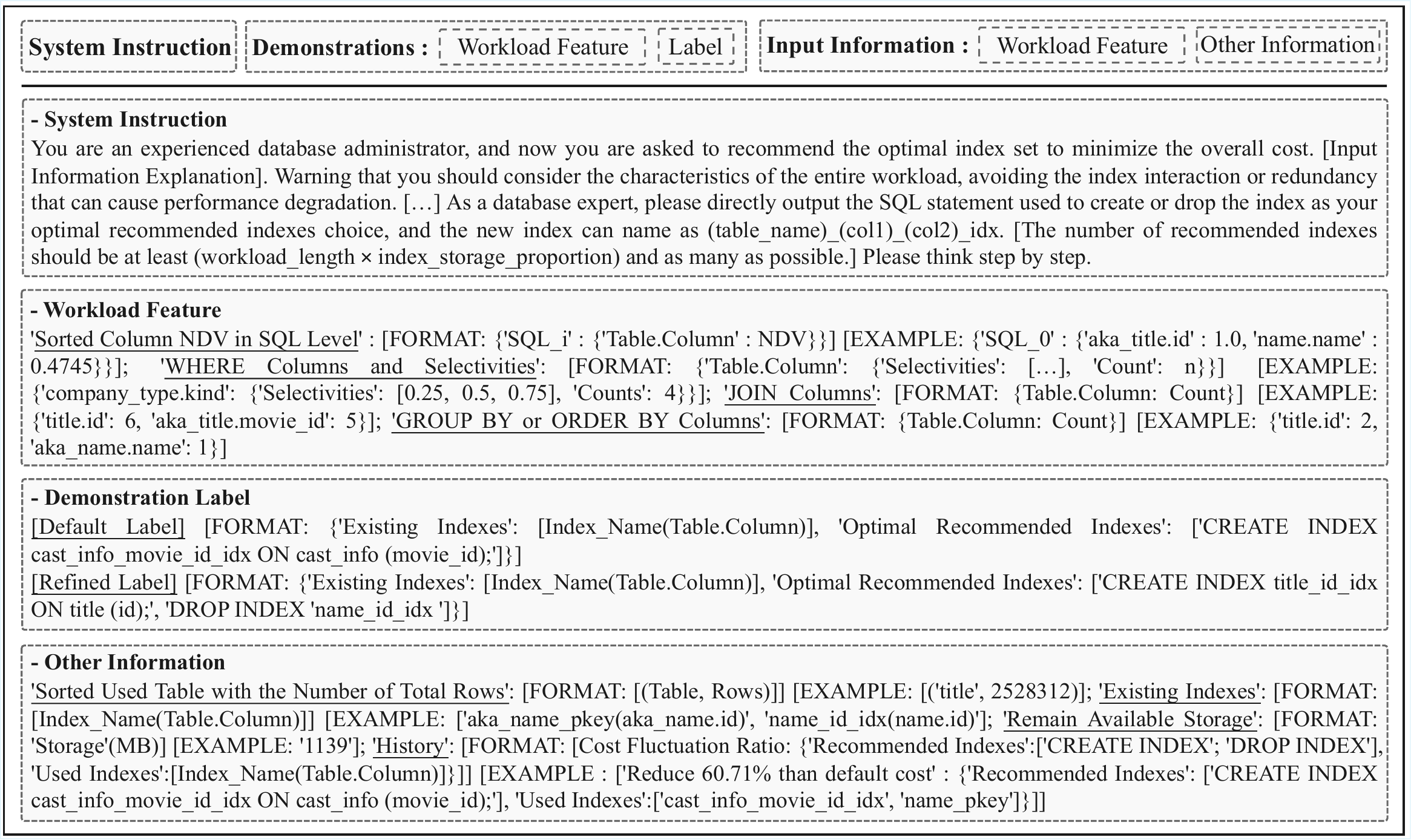}
	\caption{\label{fig:sysprompt} Illustration of \model's prompt, including its overall structure and specific content.}
\end{figure*}
}

\newcommand{\figsqlpromptapp}{
\begin{figure*}[h]
    \centering
    \includegraphics[width=0.8\textwidth]{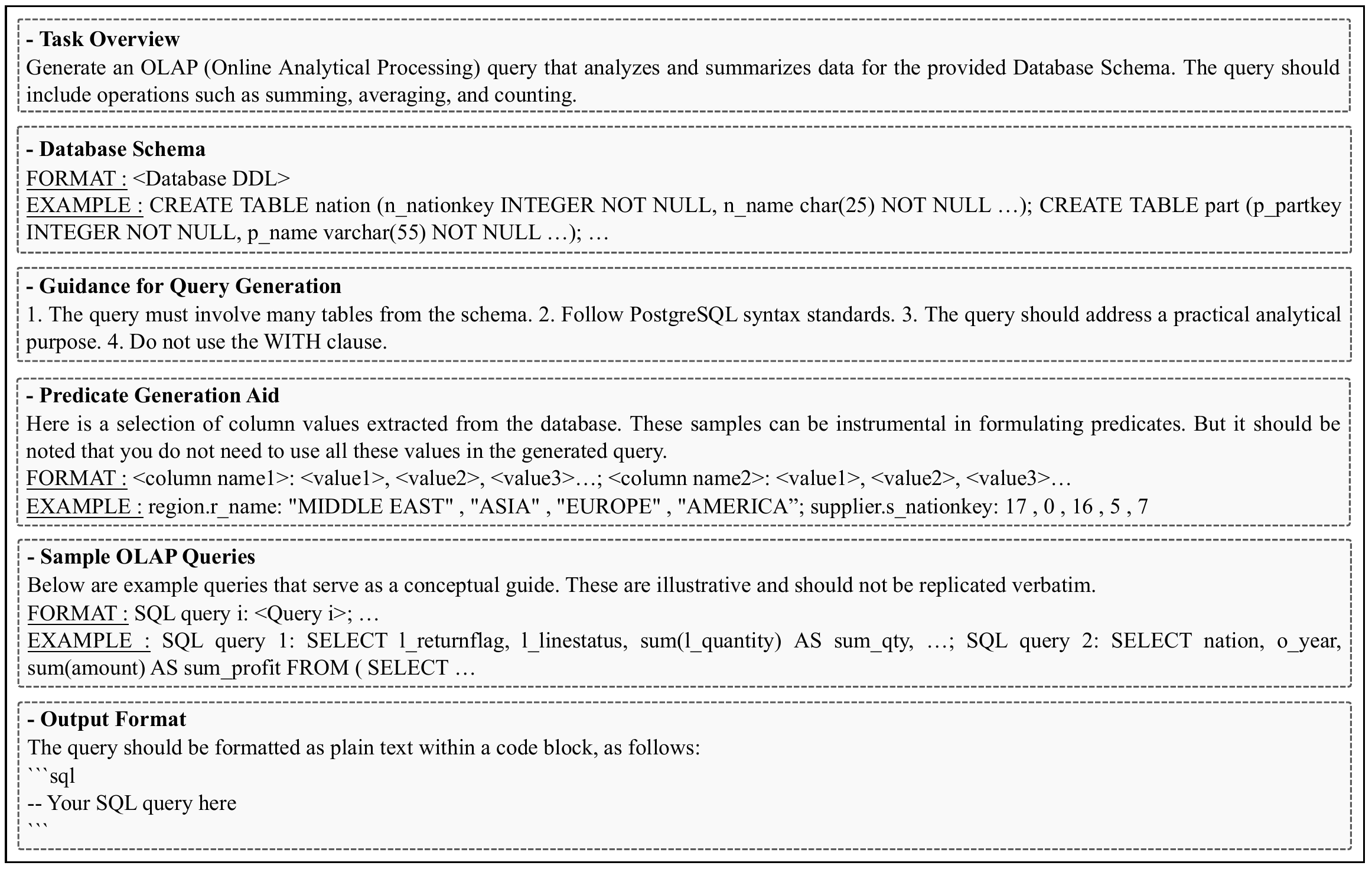}
	\caption{\label{fig:sqlprompt} Prompt for SQL synthesis.}
\end{figure*}
}

\newcommand{\figtpcdsall}{
\begin{figure*}[t]
    \centering
    \includegraphics[width=0.82\textwidth]{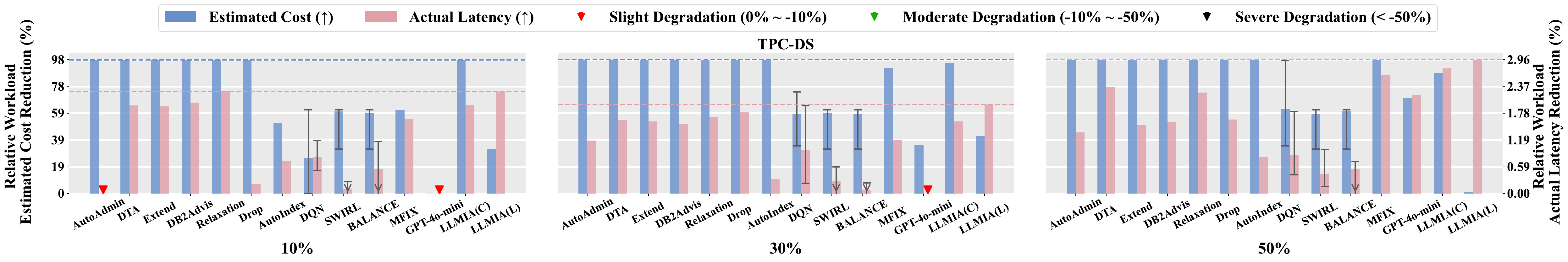}
	\caption{\label{fig:tpcdsall} Performance evaluation for TPC-DS without queries deletion.}
\end{figure*}
}

\section{INTRODUCTION}\label{sec:intro}
\label{sec:intro}
Index recommendation aims to identify optimal indexes under constraints (e.g., storage budget) based on the workload to optimize the performance of database management systems (DBMSs). 
Traditionally, database administrators (DBAs) manually recommend indexes based on their expertise. However, index recommendation is an NP-complete problem~\cite{npcomplete}, making manual tuning infeasible for complex workloads and large-scale database instances worldwide. To automate index recommendation, researchers have proposed various methods that can be broadly categorized as heuristic search-based approaches~\cite{magicmirror} and learning-based approaches~\cite{swirl, balance, autoindex, mfix}.
These solutions can be typically decomposed into three core components: 
Firstly, (1) \textit{Candidate Generation} constructs the initial search space for possible indexes. 
Then, (2) \textit{Index Selection} explores and refines the recommended index set, iteratively with the guidance from (3) \textit{Benefit Evaluation} for the expected performance improvement estimations.




\figlc

\vpara{Limitations of Existing Methods.} 
Although existing methods can recommend appropriate indexes with minimal human intervention, their real-world performance is limited by a key pitfall: they rely heavily on the estimated costs rather than actual latencies to guide index selection. This reliance exists because these methods require frequent benefit evaluations throughout the recommendation process. In heuristic-based approaches~\cite{DTA, Extend, AutoAdmin, drop, relaxation}, exploring the vast search space—caused by numerous columns in complex database schemas—demands extensive benefit evaluations. In our preliminary studies, heuristics such as AutoAdmin~\cite{AutoAdmin}, DTA~\cite{DTA}, Relaxation~\cite{relaxation}, and Drop~\cite{drop} require over 20,000 benefit evaluations on TPC-DS benchmark. Similarly, learning-based approaches~\cite{swirl, dqn, balance, mfix} employ online reinforcement learning or Bayesian optimization algorithms to iteratively refine recommendations from a discrete set of index candidates. These methods require 100$\sim$400 iterations, with each iteration involving at least one benefit evaluation. Since index creations and actual workload executions are time- and resource-intensive, most methods~\cite{magicmirror, swirl, balance} rely on the estimated costs to efficiently evaluate index benefits. Specifically, they utilize the \texttt{what-if} caller~\cite{whatif}, which enables virtual index management without modifying the database, with the \texttt{EXPLAIN} cost estimations to approximate the candidate indexes' benefits. However, there remains a significant gap between estimated costs and actual latencies. As shown in Figure~\ref{fig:lc}, we plot the estimated costs alongside the actual latencies on TPC-H benchmark under various index sets, which indicates that estimated costs are not always proportional to actual latencies. Given that existing methods rely heavily on these estimations for index recommendation, such discrepancies can cause severe reward hacking, ultimately harming the real-world performance of the recommended indexes.

\vpara{Motivation.}
Unlike exhaustive search in a vast space, experienced DBAs identify appropriate indexes based on a few interactions with the DBMS. For high-priority workloads, they involve actual execution—building indexes and running the workload—to obtain accurate feedback, and DBAs could find suitable indexes within about 10 iterations, making the overhead acceptable for these critical workloads. This efficiency coupled with satisfactory performance stems from two key abilities: (1) extensive experience in index recommendation, enabling DBAs to initialize strong indexes; (2) iterative refinement of recommendations based on the DBMS's feedback. 

Meanwhile, recent advances in generative large language models (LLMs) such as GPT-4\cite{gpt4-turbo} and DeepSeek\cite{deepseek} have transformed the natural language processing (NLP) field. Pre-trained on massive corpora, LLMs exhibit powerful human-like reasoning abilities and have recently been used for complex reasoning tasks such as mathematical problem solving\cite{math}, code generation\cite{swebench, codeforces}, and scientific research~\cite{deepresearch}. Motivated by these successes, we explore the feasibility of using LLMs to emulate the role of experienced DBAs and perform index recommendations. While LLMs have demonstrated strong general problem-solving capabilities, their potential to act as automated DBAs and provide efficient, effective index recommendations has received little attention to date.

\vpara{Challenges of Leveraging LLMs for Index Recommendation.} Preliminary experiments indicate that simply providing the workload to an LLM may result in suboptimal index recommendations (see Section~\ref{sec:main_results}), which is likely due to the scarcity of index recommendation-related data during LLMs' training process~\cite{domain0,domain1}. Thus, the first key challenge is: How can we inject database expertise into LLMs to make them more suitable for index recommendation (\textbf{C1})?
Then, providing only superficial workload features—such as column names—as input for LLMs remains insufficient. Experienced DBAs rely on various detailed workload and database characteristics, such as row counts, data types, and the number of distinct values (NDV), to guide their decisions. To narrow the gap between input and the DBAs' decision-making process, and to better align LLM behavior with DBA practices, it is crucial to make these informative features explicitly available to the LLM. Thus, our second challenge is: how can we comprehensively extract and provide these critical workload features to LLMs (\textbf{C2})?
As noted above, experienced DBAs rely on a few feedback from the DBMS to iteratively refine their recommendations. For LLM-based index recommendation, the third challenge is: How can we incorporate database feedback into the LLM’s recommendation process (\textbf{C3})?

\vpara{Our Proposal.}
Facing these challenges, we propose \model, an \underline{LLM}-based \underline{I}ndex \underline{A}dvisor that enables flexible and effective index recommendation. The core idea is to leverage the LLMs' in-context learning capabilities with informative demonstration construction and comprehensive workload features integration, while further improving recommendations via database feedback-driven index selection and refinement.

First, to inject database expertise into LLMs, we employ few-shot in-context learning guided by carefully selected demonstrations, rather than model fine-tuning. Fine-tuning typically requires substantial computational resources and large-scale training datasets. When only limited training data is available, fine-tuning LLMs can lead to overfitting, which in turn limits the model’s ability to generalize to new scenarios such as unseen workloads, database instances, or different hardware environments. In contrast, few-shot in-context learning leverages a compact but high-quality demonstration pool and effective retrieval strategies, enabling knowledge injection without altering the LLM. 
In this work, we propose a novel approach for constructing the demonstration pool. To ensure its diversity and encompass a broad spectrum of scenarios, we utilize GPT-4-Turbo~\cite{gpt4-turbo} to synthesize a variety of database workloads, and then label index refinement demonstrations using an ensemble of heuristic-based methods. Each demonstration consists of a quadruple \textless database, workload, current indexes, refinement action\textgreater, where action specifies the operations (e.g., CREATE or DROP INDEX) required to refine the current indexes. Once the pool is constructed, it can be used to provide guidance for a new workload. During online recommendation, given a workload, \smodel performs similarity ranking to select the most relevant demonstrations from the pool, thereby effectively injecting database expertise into the LLM’s context and addressing challenge \textbf{C1}.




Second, leveraging the flexibility of the LLM to process diverse input features in text format, we design a comprehensive feature extraction module that incorporates information from columns, predicates, key operations (such as JOIN and GROUP BY), and database statistics obtained from the DBMS. This provides informative workload features to the LLM for more effective index recommendation, addressing \textbf{C2}.


Third, we treat online index recommendation as an iterative process that leverages database feedback in two stages: index selection and index refinement. Starting from an initial index set—typically empty—our method proceeds iteratively. In each iteration, the LLM generates multiple candidate refinement actions that could potentially improve the current index set. Each candidate action is applied to the current index set for evaluation, and the best-performing set is selected based on database feedback, enabling feedback-guided index selection. The selected index set and its feedback are then provided to the LLM in the next iteration for further refinement, enabling feedback-guided index refinement. This process repeats until reaching the maximum iterations, thus addressing \textbf{C3}.

\vpara{Contributions.}
Our main contributions are as follows:

\begin{itemize}[leftmargin=1em]
\setlength\itemsep{0em}
    \item Inspired by the behavior of experienced DBAs, we propose \model, an LLM-based index advisor that can identify appropriate indexes for new workloads with only a few interactions with the DBMS. This represents a shift from inefficient exhaustive search to a new paradigm leveraging LLMs' semantic understanding and reasoning capabilities for index recommendation, distinguished from existing methods.
    \item To achieve this, we introduce a novel demonstration pool construction strategy to inject database expertise into the LLM, a comprehensive workload feature extractor to provide rich input features, and a feedback-guided index recommendation framework that enables the LLM to iteratively improve recommendations based on the database feedback.
    \item We conduct extensive evaluations of \smodel against 12 baselines on five widely used benchmarks, demonstrating it can recommend superior indexes with minimal database feedback. We further verify \smodel on 2 real-world benchmarks, highlighting its out-of-the-box adaptability to unseen workloads, database instances, and hardware environments. Finally, comprehensive ablation studies validate the effectiveness of each component. Our code is publicly available.\footnote{We release our code at: \url{https://github.com/XinxinZhao798/LLMIndexAdvisor}}
\end{itemize}

\section{RELATED WORK}

\subsection{Index Recommendation}
We categorize the existing index recommendation studies into three main types: heuristic methods, reinforcement learning (RL)-based methods, and other methods~\cite{breakingitdown}.

\vpara{Heuristic methods.}
Heuristic methods~\cite{magicmirror} greedily explore index candidates guided by estimated costs. Depending on the initial index set, they follow either (1) a \textit{bottom-up} strategy~\cite{DTA,Extend,AutoAdmin} iteratively adding indexes to an empty set, or (2) a \textit{top-down} strategy~\cite{drop, relaxation} iteratively removing indexes from a large initial set. Although some rules such as splitting indexes into shared and residual columns or removing redundant columns~\cite{relaxation} can refine the search space, these methods still require numerous iterations with substantial online recommendation time when facing more sophisticated database schemas. Moreover, their iterative decisions rely heavily on cost estimation, whose inaccuracies can mislead the search.

\vpara{RL-based methods.}
Given the workload feature as the state representation, RL-based methods~\cite{swirl, dqn, balance} select an action from the action space constructed by index candidates through the policy model. After updating the current state through simulating the selected action via \texttt{what-if} caller, the benefit could be calculated to guide the critic model. This process iteratively refines both the policy model and the critic model, improving the decision-making process over time. 
However, RL-based methods often require hundreds of online iterations to achieve satisfactory performance. Although existing studies could utilize an offline-trained model for single-iteration inference to reduce online time, they may suffer from significant performance degradation.

\vpara{Other methods.}
AutoIndex~\cite{autoindex} utilizes Monte Carlo Tree Search to make incremental recommendation based on existing indexes. MFIX~\cite{mfix} employs Bayesian Optimization approach~\cite{bo}, which leverages an optimal balance between exploitation and exploration. However, these methods still face similar challenge of numerous benefit evaluations.

To address these challenges, we propose an out-of-the-box, tuning-free index advisor that leverages large language models (LLMs) to achieve both efficiency and efficacy. By equipping the LLM with DBA expertise through demonstration-based in-context learning and incorporating database feedback, our method enables rapid and accurate index recommendation within just a few interactions with the DBMS.


\subsection{Large Language Models for Databases}
Recently LLMs have gained prominence for their extraordinary performance across various tasks, and growing research has emerged to explore LLMs' applications in DBMSs, such as text-to-SQL~\cite{resdsql, codes, dinsql}, knob tuning~\cite{llmtune, gptuner}, and database diagnosis~\cite{dbot, dbgpt}. However, these tasks involve distinct feature extraction and task-specific designs, making it challenging to directly apply existing methods for index recommendation. 
IdxL~\cite{idxl} introduces a query-level index recommendation approach that fine-tunes the T5 language model~\cite{t5} using massive $<$SQL query, appropriate indexes$>$ training pairs. However, except the resource demands of fine-tuning, IdxL struggles with workload-level index recommendation because it fails to capture relationships across multiple SQL queries in a workload. Moreover, models trained via fine-tuning often struggle to generalize to new environments due to distribution shifts between training and deployment settings.

In contrast, \smodel offers a tuning-free, workload-level index advisor that leverages the strong few-shot in-context learning abilities of LLMs, making it an out-of-the-box solution for database index recommendation.



\section{\protect\smodel OVERVIEW}
In this section, we define the index recommendation problem with relevant preliminaries (Section~\ref{problem}), and present an overview of our proposed \smodel (Section~\ref{overview}).

\figoverview





\subsection{Problem Formulation}

\label{problem}



Index recommendation denotes identifying optimal indexes for a given workload while satisfying constraints like the storage budget of indexes. 

\vspace{0.05in}\noindent \textit{Definition. Index Recommendation Problem (IRP).}
Given a workload \(W = \{q_1, ..., q_m\}\) with \(m\) SQL queries, we extract all columns referenced in \(W\) to generate a set of candidate indexes \(\mathcal{I} = \{i_1, i_2, ..., i_n\}\), encompassing all possible single- and multi-column indexes. Under a storage constraint \(S_c\), the Index Recommendation Problem (IRP) seeks an optimal subset \(I^{*} \subseteq \mathcal{I}\) that minimizes the total workload cost \(C\):
\begin{equation}
\begin{aligned}
    I^{*} = \arg\min_{I \subseteq \mathcal{I}} C(W, I)\,, \\
    \text{s.t. } S(I) \leq S_c\,.
\end{aligned}
\end{equation}
Here, \(S(I) = \sum_{i\in I} s_{i}\) is the total storage consumption of the selected indexes \(I\), where \(s_{i}\) denotes the storage of index \(i\). The workload cost is defined as \(C(W, I) = \sum_{j=1}^{m} \text{cost}(q_{j}, I)\), where \(\text{cost}(\cdot)\) can represent either estimated costs or actual latencies. Estimated costs are typically obtained using \texttt{EXPLAIN} and the \texttt{what-if} caller to simulate index effects; actual latencies are measured by physically building the indexes and executing the workload. Traditional methods generally rely on estimated costs because the cost calculation must be performed frequently. In contrast, since \smodel requires much fewer interactions with the DBMS, using actual latencies is practical and acceptable for important or recurring workloads. Of course, \smodel can also use estimated costs if preferred.

\subsection{System Overview}
\label{overview}
Figure~\ref{fig:overview} presents an overview of \model, which consists of two stages. The first stage constructs the demonstration pool offline to capture database expertise. The second stage describes the index recommendation framework, combining prompt construction with an LLM-driven iterative refinement to emulate the decision-making of DBAs.

\vpara{Stage 1: Demonstration Construction (Section~\ref{sec:demo}).}
Each demonstration in our approach consists of a quadruple \textless database, workload, current indexes, refinement action\textgreater, illustrating the action needed to refine a set of current indexes for improved performance on a given database and workload. To build a high-quality and diverse demonstration pool, we offline synthesize workloads and flexibly generate \textless current indexes, refinement action\textgreater\ pairs. Based on whether the current indexes are empty or non-empty, we define two types of demonstrations: initial refinement demonstrations and incremental refinement demonstrations.

\vpara{Stage 2: Index Recommendation Pipeline (Section~\ref{sec:indexrecom}).}
To unlock the potential of LLMs for index recommendation, we use a tuning-free LLM with in-context learning (ICL). Specifically, for a given target workload, we construct the input prompt in three parts: (1) system instruction, (2) dynamically matched demonstrations, and (3) detailed input information. The index recommendation process begins with an empty index set, and the LLM iteratively generates actions to refine the indexes. At each iteration, we incorporate database feedback to guide candidate index selection and update the feedback in the prompt, allowing the LLM to adjust based on the database’s responses for improved refinement. The iteration process stops once the maximum number of steps is reached, and the best-found index set is then returned to the user.

\section{Demonstration Construction}
\label{sec:demo}
To enhance the LLM’s database expertise, we leverage in-context learning (ICL) to inject expert knowledge via task-related demonstrations. Therefore, our first goal is to construct a high-quality demonstration pool covering diverse scenarios.

\subsection{Workload Generation}
\label{sec:workload_gen}
Although existing benchmarks like TPC-H and TPC-DS provide SQL templates for workload generation, we do not use them for two reasons. First, relying solely on templates limits the diversity of workloads. 
Second, template-based workload generation leads to high similarity with the queries in test workloads, which the LLM could imitate the demonstrations' actions, thereby yielding unconvincing results.
Therefore, to build a robust demonstration pool, we additionally synthesize new SQL queries for these database schemas. 



Given that OLTP benchmarks are primarily used to emulate concurrent transactions in commercial environments through repeating simple queries, preliminary attempts show that index recommendation is easier in these settings due to the small search space (e.g., the Twitter database in OLTP-Bench\cite{oltpbench} averages only two candidate columns per table). Therefore, following prior work~\cite{magicmirror, balance, swirl, mfix}, this paper focuses on index recommendation for OLAP benchmarks, i.e., the workloads containing very complex analytical queries.
As SQL queries often involve multiple operations such as filters and joins, we follow~\cite{llmtune} and use GPT-4-Turbo~\cite{gpt4-turbo} to synthesize complex SQL queries. The prompt provided to GPT-4-Turbo includes (1) a task overview, (2) the database DDL schema, (3) constraint conditions, (4) example SQL queries, and (5) the required output format. The constraints are as follows:
\begin{itemize}[leftmargin=1em]
    \item \textbf{Complexity:} Queries must involve multiple tables.
    \item \textbf{Syntactical Integrity:} Queries should follow specific syntax standards (PostgreSQL in this paper).
    \item \textbf{Semantic Validity:} Sampled real values from database aim to ensure that queries serve meaningful analytical purposes.
\end{itemize}

We explicitly instruct GPT-4-Turbo to use the provided SQL examples as references only, discouraging direct copying or simple modification to avoid generating queries overly similar to existing benchmarks.
After generation, we apply post-processing to ensure the validity and diversity. For validity, we run the \texttt{EXPLAIN} command to detect syntax errors, using GPT-4-Turbo’s reflection to resolve any issues; queries that remain erroneous are discarded. For diversity, we randomly sample tables, column values, and example queries in each prompt and set the sampling temperature to 1.0 to encourage varied outputs. Finally, new workloads are formed by randomly selecting from the pool of valid synthetic SQL queries.

\figdc

\subsection{Refinement Demonstration Annotation}
To enable LLMs to learn how to refine index sets for better performance, we annotate \textless current indexes, refinement action\textgreater\ pairs for each synthetic workload. These annotated pairs could serve as demonstrations to provide database expertise in the LLM’s context. There are typically two types of current index states: empty and non-empty. When the current indexes set is empty, we call this an initial refinement demonstration; otherwise, it is an incremental refinement demonstration.

To automatically generate these demonstrations, we use the following pipeline: For each synthetic workload, several existing methods are employed to recommend multiple candidate index sets. To improve annotation efficiency, we rank these candidates by estimated costs (using \texttt{what-if} caller and \texttt{EXPLAIN}) rather than actual latencies. The best-performing set of indexes can be used to form the initial refinement demonstrations: here, the refinement action involves the CREATE INDEX statements needed to build the recommended set from an empty index set. For incremental refinement demonstrations, we treat each sub-optimal candidate as the current index set, and define the refinement action as the set of CREATE and DROP INDEX operations required to transform it into the best-performing set. Initial refinement demonstrations help LLMs recommend indexes starting from scratch, while incremental refinement demonstrations enable iterative, self-refining index management with existing indexes. 

We illustrate an example of constructing two types of demonstrations in Figure~\ref{fig:dc}. In practice, we implement several heuristic-based methods to recommend index candidates, including AutoAdmin\cite{AutoAdmin}, Extend\cite{Extend}, Drop\cite{drop}, Relaxation~\cite{relaxation}, DTA~\cite{DTA}, and DB2Advis\cite{db2advis}. We do not use learning-based methods here because they require time- and resource-intensive iterative model training. 
When facing new database instances, we do not need to regenerate workloads or re-annotate demonstrations—the offline demonstration pool already enables the LLM to learn index recommendation knowledge and generalize to unseen environments, providing the out-of-the-box capability (see Section~\ref{sec:main_results} for details).

\section{Index Recommendation}
\label{sec:indexrecom}
Given a workload, database, and storage constraint, \smodel first constructs an input prompt containing core information for index recommendation, including workload features (Section~\ref{sec:wf}), retrieved demonstrations (Section~\ref{sec:icl}), system instructions, and other relevant details (Section~\ref{sec:pe}). It then applies an LLM-driven iterative refinement to identify suitable indexes (Section~\ref{sec:inferscaling}). Starting from an initial index set (typically empty), the LLM generates multiple candidate actions to refine the index set, and the best-performing set is selected for the next iteration. After each iteration, the prompt will be updated to accommodate the latest database feedback.


\subsection{LLM Input}

\subsubsection{Workload Feature}
\label{sec:wf}




Existing methods typically consider solely the columns present in the workload as features, which is insufficient for LLMs. In contrast, human experts and LLMs can leverage a broader set of heterogeneous features—such as SQL predicates and column-specific statistics—for more accurate index recommendation. Since the main goal of indexes is to reduce time-consuming full table scans and accelerate data retrieval, especially for large tables, we follow the principles outlined below to extract workload features:


\begin{itemize}[leftmargin=1em]
    \item \textbf{Principle 1:} Columns appeared in \textmd{WHERE} predicates tend to be candidates for index construction, especially preferring those that retain fewer rows after conditional filtering.
    \item \textbf{Principle 2:} Columns in \textmd{JOIN}, \textmd{GROUP BY}, or \textmd{ORDER BY} conditions are typically regarded as potential candidates that can help avoid full table scans for value retrieval or sorting.
    \item \textbf{Principle 3:} Columns with a higher number of distinct values (NDV) are more suitable for index creation, as their uniqueness of data value tends to directly index target rows.
\end{itemize}

To accurately obtain the above workload details, we implement a comprehensive feature extraction mechanism capable of handling sophisticated analytical SQL queries, which can extract different conditional predicates from the SQL statement. Specifically, the extracted features are as follows:

\begin{itemize}[leftmargin=1em]
    \item \textbf{Used Column Information in SQL-Level:} Columns appeared in each SQL of the workload, along with their corresponding NDV, number of rows, and data type. The specific information of the columns can be retrieved from the database statistics before online recommendation.
    \item \textbf{``WHERE'' Predicates and Corresponding Selectivity:} All \textmd{WHERE} predicates appeared in the workload, along with their selectivity, which is defined as the ratio of rows that satisfy the condition to the total number of rows. 
    \item \textbf{``JOIN'', ``GROUP BY'', and ``ORDER BY'' Related Columns:} Columns appeared in all \textmd{JOIN}, \textmd{GROUP BY} and \textmd{ORDER BY} conditions of the workload, along with their frequencies of occurrence.
\end{itemize}

\subsubsection{In-Context Learning}
\label{sec:icl}

Considering the limited database expertise of general LLMs, we employ few-shot in-context learning (ICL)~\cite{icl, icl0, icl1} to inject domain-specific knowledge into the LLM's context, thereby enabling the LLM to efficiently adapt to new tasks without extensive fine-tuning.

Given a demonstration pool, dynamically retrieving or matching the most relevant demonstrations remains an active research topic in NLP~\cite{icl-survey-man-2024}. To enable similarity computation between a new workload and those in the demonstration pool, we consider two key features: the frequency of column occurrences within the workload and their corresponding numbers of distinct values (NDV). Both features are min–max normalized within the workload, and the resulting pairs are sorted in descending order to construct the final workload meta-feature. Each workload is thus represented as $[(f_1, ndv_1), (f_2, ndv_2), ..., (f_m, ndv_m)]$, where $m$ denotes the number of columns involved.
We explore the following three strategies for demonstration matching:
\begin{itemize}[leftmargin=1em]
    \item \textbf{Random Sample:} Randomly select demonstrations from the pool.
    \item \textbf{Cosine Similarity Ranking:} Compute the cosine similarity between the meta-feature of the target workload and those of all workloads in the pool, then rank demonstrations in descending order of similarity. When the workloads involve different numbers of columns, we trim the longer one to match dimensions before similarity computation.
    \item \textbf{K-Means Clustering~\cite{kmeans4demos0}:} Use the k-means clustering algorithm~\cite{kmeans} to identify $k$ cluster centers, select the matching cluster based on the Euclidean distance between cluster centers and the input workload's meta-feature, then randomly sample demonstrations from the matched cluster.
\end{itemize}

Based on the experimental results presented in Section~\ref{sec:abla4demomatch}, cosine similarity ranking is adopted as the final demonstration-matching strategy. The top two demonstrations are injected into the LLM’s input prompt and dynamically updated according to their ranking throughout the iterative refinement process (see Section~\ref{sec:inferscaling} for details).



\subsubsection{Prompt Construction}
\label{sec:pe}
The prompt\footnote{The prompt overview of \smodel is available at: \url{https://github.com/XinxinZhao798/LLMIndexAdvisor/blob/main/illustrations/LLMIA_Input_Info.png}.} of \smodel is primarily composed of the following three components:

\begin{itemize}[leftmargin=1em]
    \item \textbf{System Instruction} includes the task overview, specific description of input information, output format, as well as some simple suggestions like recommended indexes' order. 
    \item \textbf{Demonstrations} are selected from the demonstration pool using the strategy outlined in Section~\ref{sec:icl}. 
    \item \textbf{Input Information} consists of workload features described in Section~\ref{sec:wf}, along with additional details such as current indexes, available storage, and historical information. The historical information comprises database feedback (e.g., estimated cost or actual latency) and the indexes used in previous query plans. These elements are updated continuously throughout the refinement process.
\end{itemize}


\subsection{LLM Index Recommendation with Database Feedback}
\label{sec:inferscaling}
Similar to bottom-up heuristic methods, we frame \model's index recommendation as a greedy search process. However, unlike existing methods that enumerate all possible actions at each iteration (resulting in a prohibitively large search space), we employ an LLM to generate potentially useful actions, guided by both demonstrations and database feedback, making the search process significantly more efficient. Consequently, \smodel implements a “self-refinement” mechanism through multi-iteration inference, interacting with the DBMS to iteratively improve its recommendations.

Specifically, given a database $D$, a workload $W$, and an initial index set $I_{0}$ (typically empty), \smodel proceeds iteratively. At each iteration $t$, the LLM produces $K$ potentially useful actions to refine the current index set. Formally,
\begin{equation}
\begin{aligned}
    LLM(\text{Ins},\, \text{Demo}(W),\, \text{Info}(W, D, I_{t}))
    \rightarrow \{a_t^{0}, \ldots, a_t^{K}\}
\end{aligned}
\end{equation}
\noindent where \text{Ins} denotes a fixed system instruction, \text{Demo($\cdot$)} represents the retrieved demonstrations, and \text{Info($\cdot$)} captures fine-grained input information. $I_{t}$ is the current index set, and $\{a_t^{0}, ..., a_t^{K}\}$ are the $K$ actions generated by the LLM at this iteration. To encourage diversity in the actions, we set the LLM sampling temperature to 0.8.



We then adopt a major voting strategy to aggregate the most common indexes among the $K$ actions into a new action $a_t^{agg}$. Specifically, we gather all actions $\{a_t^{0}, ..., a_t^{K}\}$ and sort the \textmd{``CREATE INDEX''} and \textmd{``DROP INDEX''} statements by frequency. 
To construct the aggregated action, we include all ``DROP INDEX'' statements with more than one recommendation.
For "CREATE INDEX" statements, we prioritize single-column indexes to minimize storage usage. 
Since composite indexes could support multiple attributes indexing to accelerate queries with predicates on both columns, we include multi-column indexes only if they receive multiple votes.


We include $a_t^{agg}$ among the candidate actions because this consensus-driven action, formed via major voting, often captures the collective judgment (self-consistency) of the LLM and thus has a higher chance of outperforming individual candidates~\cite{selfconsistency, sc0}. By applying all candidate actions $\{a_t^{0}, …, a_t^{K}, a_t^{agg}\}$ to $I_{t}$, we obtain $K+1$ new index sets $\{I_{t+1}^{0}, …, I_{t+1}^{K}, I_{t+1}^{K+1}\}$. We then adopt a “Best-of-N” strategy, evaluating each candidate (using either estimated cost or actual latency) and selecting the best-performing set as $I_{t+1}$ for the next iteration  (feedback-guided index selection).

Before the next iteration, we update the LLM’s prompt, including both the demonstrations and the input information. For demonstrations, we replace the current examples with top-ranked new ones according to the previous similarity-based ranking, skipping those already seen, to provide diverse references and balance exploration and exploitation for the LLM. We use initial refinement demonstrations if the current index set is empty and incremental refinement demonstrations otherwise. In addition, the input information—including the current indexes, available storage, database feedback, and indexes used—is  recalculated after each iteration to provide the LLM with timely feedback and enable better recommendations (feedback-guided index refinement).

This optimization proceeds until no further performance improvement is observed or the maximum number of iterations is reached, at which point the best index set found during the process is selected as the final recommendation.

\section{Experiments}
In this section, we conduct comprehensive experiments to evaluate the performance of our proposed \model, answering the following questions:
\begin{itemize}[leftmargin=1em]
    \item \textbf{RQ1:} How does \smodel improve the workload performance compared with the existing methods across various database schemas and storage constraints?
    \item \textbf{RQ2:} How does the cost, in terms of efficiency and overhead (i.e., LLM API expense), incurred by our \smodel compare with that of existing methods during online recommendation?
    \item \textbf{RQ3:} Considering that \smodel consists of multiple components, how do they enhance the overall performance of the index recommendation pipeline?
    \item \textbf{RQ4:} How about the additional overheads of our \smodel during offline preparation?
    
\end{itemize}

\subsection{Experimental Settings}

\subsubsection{Environments}
We perform all experiments on PostgreSQL 12.2 database system, hosted on a server equipped with an Intel(R) Xeon(R) CPU E5-2650 v4 @ 2.20GHz featuring 12 cores and 24 threads, along with 64GB of RAM. 
To support the benefit estimation, we implement the \texttt{what-if} caller using HypoPG extension~\cite{whatif} to simulate index creation or deletion, and obtain the SQL query's estimated cost under virtual indexes via executing the \texttt{EXPLAIN} command.

\begin{table*}[t]

    \caption{Database information and workload statistics. We present the ``min / max / avg'' value of each clause in the test set and constructed demonstrations. Note that the SSB and real-world benchmarks include only the test workloads' statistics.}
    \centering
    \vspace{-0.7em}
    \begin{adjustbox}{width=0.86\textwidth}
    \label{tab:data}

    \begin{tabular}{@{}l|c|c|c|w{c}{1cm}w{c}{1cm}|w{c}{1.6cm}w{c}{1.6cm}|w{c}{1.6cm}w{c}{1.6cm}|w{c}{2.2cm}w{c}{2.2cm}@{}}
    
    \toprule
    \multirow{2}{*}{\textbf{Database}} & \multirow{2}{*}{\textbf{Size}} & \multirow{2}{*}{\textbf{\# Tables}} & \multirow{2}{*}{\textbf{\# Demo.}} & \multicolumn{2}{c}{\textbf{\# Queries}} & \multicolumn{2}{c}{\textbf{\# WHERE Predicates per SQL}} & \multicolumn{2}{c}{\textbf{\# JOIN Predicates per SQL}} & \multicolumn{2}{c}{\textbf{\# GROUP BY / ORDER BY Columns per SQL}} \\
    \cmidrule(r){5-6}\cmidrule(r){7-8}\cmidrule(r){9-10}\cmidrule(r){11-12} 
    &  &  &  & \textbf{Bench} & \textbf{Demo.} & \textbf{Bench} & \textbf{Demo.} & \textbf{Bench} & \textbf{Demo.} & \textbf{Bench} & \textbf{Demo.} \\
    \midrule
    TPC-H~\cite{tpch} & 7.2GB / 101GB & 8 & 192 & 19 & 745 & 1 / 4 / 2.11 & 0 / 8 / 1.47 & 0 / 7 / 2.87 & 2 / 13 / 4.11 & 0 / 7 / 1.95 & 2 / 7 / 2.06 \\
    JOB~\cite{job} & 6.9GB & 21 & 198 & 113 & 950 & 1 / 14 / 1.72 & 0 / 8 / 1.66 & 5 / 24 / 11.84 & 1 / 12 / 3.28 & 0 / 0 / 0.0 & 0 / 6 / 1.52 \\
    TPC-DS~\cite{tpcds} & 2.3GB & 24 & 200 &  90 & 1003 & 1 / 14 / 1.98 & 0 / 25 / 2.34 & 0 / 21 / 6.13 & 0 / 14 / 3.2 & 0 / 17 / 4.88 & 0 / 15 / 3.74 \\
    SSB~\cite{ssb} & 8.0GB & 5 & - & 13 & - & 1 / 4 / 2.92 & - & 1 / 4 / 2.76 & - & 0 / 6 / 3.68 & - \\
    SSAG & 58GB & 13 & - & 6 & - & 3 / 8 / 4.5 & - & 0 / 2 / 0.83 & - & 0 / 5 / 2.33 & - \\
    AMPS & 14GB &  6 & - & 95 & - & 0 / 3 / 0.77 & - & 0 / 1 / 0.03 & - & 0 / 1 / 0.16 & - \\
    
    \bottomrule
        
    \end{tabular}

    \end{adjustbox}

\end{table*}

\subsubsection{Benchmarks and Datasets}
We evaluate five standard OLAP benchmarks with complex analytical SQL queries, including TPC-H~\cite{tpch}, JOB~\cite{job}, TPC-DS~\cite{tpcds}, and SSB~\cite{ssb}.
The JOB benchmark involves 113 query templates based on the Internet Movie Database, while TPC-H, TPC-DS, and SSB benchmarks involve 24, 99, and 13 query templates respectively with synthetic database. For TPC-H benchmark, we conduct evaluations on two databases of different sizes, setting the scale factor (SF) as 5 and 50 respectively.
Since we focus on the workload-level index recommendation, referring to prior studies~\cite{magicmirror, balance, mfix, swirl}, we exclude those queries whose estimated costs or execution latencies are orders of magnitude higher than normal queries. The predominance of those specific cases renders the workload-level index recommendation shifting toward recommending indexes for those special queries, making the research problem less complex. 
Specifically, we exclude queries 4, 6, 9, 10, 11, 31, 35, 41, 74 in TPC-DS, and queries 2, 17, 20 in TPC-H. 



Since LLMs have likely encountered the queries from above well-known benchmarks during their pre-training, we conduct extended experiments on two real-world benchmarks from ByteDance: SSAG and AMPS, which should not be present in the LLMs' pre-training corpus. SSAG is an OLAP benchmark used for analyzing and managing slow SQL queries, including tasks like slow SQL identification and logical database analysis. AMPS is an OLTP benchmark used in AI platform services, including transactions related to user management, permission control, and task scheduling. All the pre-defined secondary indexes are removed before index recommendation.


We construct the demonstration pool with the database instances from TPC-H, JOB, and TPC-DS using the method in Section~\ref{sec:demo}. 
SSB, SSAG, and AMPS are used exclusively as test sets without additional demonstration construction, allowing us to assess \model's out-of-the-box generalization ability.
Table~\ref{tab:data} summarizes the statistics of the benchmarks and demonstrations, including database size, number of tables, number of demonstrations, and workload characteristics (e.g., number of distinct queries and average predicates per query). Statistical analysis shows that synthetic workloads have comparable complexity to standard benchmarks with greater diversity in query templates (see \# Queries).


\subsubsection{Baselines}
We compare both heuristic and learning-based methods with our \smodel for a comprehensive experimental evaluation. For heuristic methods which primarily utilize greedy search and its variants for index recommendation, we evaluate: (1) Extend~\cite{Extend}, (2) Relaxation~\cite{relaxation}, (3) DTA~\cite{DTA}, (4) DB2Advis~\cite{db2advis}, (5) AutoAdmin~\cite{AutoAdmin}, and (6) Drop~\cite{drop}. 
For the sake of fairness, we modify existing methods—Drop and AutoAdmin—to satisfy index recommendation under storage constraints.
For learning-based methods, we evaluate: 
(7) AutoIndex~\cite{autoindex}, which leverages Monte Carlo Tree Search for incremental index management; 
(8) DQN~\cite{dqn} and (9) SWIRL~\cite{swirl}, which utilize reinforcement learning, Deep Q-Network and Proximal Policy Optimization (PPO)~\cite{ppo}, for index recommendation; 
(10) BALANCE~\cite{balance}, which leverages reinforcement learning with an additional transfer mechanism for workload scenarios with query and frequency variation; and
(11) MFIX~\cite{mfix}, which utilizes Bayesian Optimization (BO)~\cite{bo} for index recommendation. 
We follow the default setting for all baselines. 
For reinforcement learning-based methods~\cite{dqn, balance, swirl}, we additionally conduct a grid search around the default values of commonly used hyperparameters (e.g., learning rate, batch size, and epochs). Specifically, for each method, we evaluate the recommended results of 27 different combinations of hyperparameters, and summarized the average, minimum, and maximum results for reporting.
Considering the predefined action space (i.e., index candidates) relevant to the workload, these methods require retraining before applying to a new workload, and the iterative training process can be treated as an online iterative index recommendation.
Additionally, to assess the efficacy of \model's index recommendation framework, we evaluate our LLM backbone: (12) GPT-4o-mini~\cite{gpt4omini}, which directly takes the SQL statements in the workload as input information for index recommendation.



\subsubsection{Metrics}
We evaluate index advisors from the following three perspectives.
\textbf{(1) Relative Workload Actual Latency Reduction} is defined as the proportion of reduction in workload actual latency after index creation (i.e., $1 - \frac{\text{Latency}_\text{w/ index}}{\text{Latency}_\text{w/o index}}$), which can be obtained from executing the workload. We treat it as the primary metric which evaluates the practical effectiveness of these methods, with a higher value denoting better performance.
\textbf{(2) Relative Workload Estimated Cost Reduction} measures the proportion of workload estimated cost's reduction after creating indexes (i.e., $1 - \frac{\text{Cost}_\text{w/ index}}{\text{Cost}_\text{w/o index}}$), which can be derived via executing the \texttt{EXPLAIN} command. Considering the existing methods recommend indexes with the guidance of the estimated cost, this metric is widely used to assess the method's optimization capability~\cite{magicmirror, breakingitdown, balance, swirl}, with a higher value signaling better performance.
\textbf{(3) Benefit Evaluation Count} denotes the number of obtaining the external database feedback during online recommendation, which can perform either \texttt{EXPLAIN} command for estimated cost or workload execution for latency, with a lower value representing better efficiency. Note that we have conducted a specific analysis for offline cost evaluation in Section \ref{eval_off_cost}.

\figmainresults

\subsubsection{Evaluation Settings}\label{sec:eval_settings}
We use GPT-4o-mini as \model's LLM backbone, and primarily design experiments under various storage constraints and application scenarios. 
\textbf{For storage constraint}, since the size of recommended indexes depends on its corresponding database size, we define this constraint as the percentage of the database size, which can be calculated as: $\frac{\text{Index Storage Constraint (MB)}}{\text{Database Size (MB)}} \times 100\%$. We evaluate three different storage constraint levels in the workload performance evaluation. 
Note that we conduct experiments for two real-world benchmarks under a single representative storage constraint, as these workloads' index requirements are relative fixed, thereby with no significant performance variance across different storage constraints.
\textbf{For application scenarios}, we evaluate two settings:
\textit{(1) Cost-Instruct (LLMIA(C))}: Similar to prior works, this mode uses the \texttt{what-if} caller and estimated cost as feedback for index management. To mitigate the limitations of cost inaccuracy, we set the sample size to 8 and run 4 iterations to ensure a comprehensive search and iteration.
\textit{(2) Latency-Instruct (LLMIA(L))}: This mode replaces virtual index management with actual index updates and uses observed workload latency as feedback. Given the higher overhead of real executions, we set the sample size to 4 and the number of iterations to 2 for LLMIA(L).
The sample and iteration parameters are kept constant across all experiments.

\subsection{Workload Performance Evaluation}
\label{sec:eval_main}

We conduct extensive experiments compared with 12 existing methods on five OLAP benchmarks under different storage constraints and two real-world benchmarks (RQ1).

\subsubsection{Standard Benchmark Evaluation}
\label{sec:main_results}
Experimental results are presented in Figure~\ref{fig:mainresults}, with the main findings as follows:

\textbf{\modelabb(L) demonstrates competitive efficacy during workload execution.} Across all scenarios, \modelabb(L) could stably achieve the satisfactory or second optimal actual latency compared with the existing methods, indicating the significance of the accurate feedback signals. For TPC-H and JOB benchmarks, \modelabb(L) consistently maintains the best, or even superior performance (TPC-H SF = 5 and JOB under storage constraint 50\%). For TPC-DS benchmark, \modelabb(L) exhibits the best performance under storage constraint 10\%, while keeping second optimal performance with only several seconds lag behind for other storage constraints. Notably, for SSB benchmark, \modelabb(L) surpasses all baselines across various storage constraints, even without constructing any relevant demonstrations, demonstrating \modelabb(L) can generalize to databases beyond those included in the demonstration pool.


\textbf{\modelabb(C) could maintain comparable workload latency with existing methods.} 
Experimental results illustrate that although \modelabb(C) may not always achieve the best estimated workload cost, it delivers promising workload execution latency that slightly outperforms existing methods. Unlike those methods directly identifying the recommended indexes according to the estimated cost, \smodel treats this feedback as a guidance for refining the current recommendation, thereby mitigating the direct impact of the estimated cost's inaccuracy.

\textbf{\model's elaborate designed framework significantly enhances the LLM's capability in index recommendation.} 
Compared with \modelabb, experimental results of directly using GPT-4o-mini present severe performance degradation in both cost and latency. We speculate that the LLM may struggle to accurately capture essential index-beneficial workload features. This could be attributed to the LLM's insufficient training in database expertise, and the information in SQL statements is limited, which can further emphasize the significance of \model's demonstrations injection and workload feature extraction, along with external database feedback instruction.

\textbf{Considerable discrepancy between the estimated cost and actual latency demonstrates the existing methods' infeasibility during practical application.}
Experimental results on the estimated cost exhibit superior performance of the heuristic methods (e.g., TPC-H SF = 50 under storage constraint 30\% and 50\%), while leading to degraded actual latency. This illustrates the estimated cost served as the external feedback could mislead the optimization direction.
Notably, \smodel primarily recommend indexes based on the LLM's intrinsic reasoning capability, and utilize the database feedback as the external information in place of a strict optimization constraint, thereby avoiding the reward hacking phenomenon.

\subsubsection{Real-World Benchmark Evaluation}
\label{sec:real_world_eval}
Considering the potential exposure of LLMs' pre-training process to the standard benchmarks' knowledge, we further evaluate \smodel on real-world benchmarks with the constructed demonstrations, and implement all baselines within ByteDance’s private environment. 
As the experimental results revealed nearly identical performance across different storage constraints, we ultimately adopt the storage constraint of 30\% for evaluation, with the experimental results presented in Figure~\ref{fig:rw}. 

\begin{table*}[t]
    \caption{\modelabb's total context (input + output) and overhead for a single-time index recommendation.}
    \centering
    \vspace{-0.7em}
    {
    \begin{adjustbox}{width=0.8\textwidth}
    \label{tab:overhead}

    \begin{tabular}{@{}lccccc@{}}
    
    \toprule
         Input + Output Tokens / Cost (\$) & \textbf{TPC-H (SF = 5)} & \textbf{JOB} & \textbf{TPC-DS} & \textbf{SSB} & \textbf{TPCH (SF = 50)}\\
        \midrule
        \modelabb(C) & 47,864 + 21,357 / 0.020  & 48,740 + 18,702 / 0.019 & 101,150 + 21,106 / 0.028 & 38,492 + 21,660 / 0.019 & 46,749 + 22,822 / 0.021 \\
        \modelabb(L) & 24,100 + 5,599 / 0.007  & 22,159 + 4,470 / 0.006 & 52,600 + 4,986 / 0.011 & 19,652 + 5,246 / 0.006  & 24,278 + 5,713 / 0.007 \\
        
    \bottomrule

    \end{tabular}
    \end{adjustbox}
    }

\end{table*}


Under out-of-distribution workloads with unseen hardware environments, \modelabb(L) still recommends effective indexes compared with existing methods. Moreover, \smodel requires no additional offline preparation for unseen scenarios, demonstrating its out-of-the-box applicability and practical value.

\figrealworld
Experimental results present no obvious performance variation across different methods for AMPS, possibly because it's an OLTP benchmark that is easy to recommend appropriate indexes within a small candidate columns.
Notably, experimental results on SSAG illustrate that most of the existing methods present decent improvements on workload estimated cost while leading to reverse optimizing on actual latency. 
Similar to the experimental results of TPC-H (SF = 50) in Figure~\ref{fig:mainresults}, we speculate that the deviation of the estimated cost calculated from the \texttt{EXPLAIN} command increases proportionally with the database size, leading to a significant decline in the performance of the baselines.
In contrast, \smodel with the advantage of minimal database interactions could directly take the actual latency for index recommendation, thereby exhibiting a notable workload actual latency improvement. 

\begin{table}[t]

    \caption{Benefit evaluation counts (\texttt{EXPLAIN} cost estimation or workload execution).}
    \centering
    \vspace{-0.7em}
    \begin{adjustbox}{width=0.44\textwidth}
    \label{tab:efficiency}

    \begin{tabular}{@{}lccccc@{}}
    
    \toprule
        & \textbf{TPC-H (SF = 5)} & \textbf{JOB} & \textbf{TPC-DS} & \textbf{SSB} & \textbf{TPCH (SF = 50)} \\
    \midrule
    AutoAdmin & 1,363 & 28,107 & 33,955 & 741 & 1,114 \\
    DTA & 7,890 & 2,944 & 21,946 & 3,807 & 1,538 \\
    Extend & 314 & 1,269 & 5,766 & 59 & 226 \\
    DB2Advis & 19 & 113 & 90 & 13 & 19 \\
    Relaxation & 474 & 7,294 & 36,809 & 308 & 176 \\
    Drop & 1,277 & 2,737 & 28,427 & 376 & 1,293 \\
    \midrule
    AutoIndex & \underline{15}  & 119  & 132 & \underline{12}  & 70 \\
    DQN & 400  & 400  & 400 & 400 & 400 \\
    SWIRL & 200  & 200  & 200 & 200 & 200 \\
    BALANCE & 200  & 200  & 200 & 200 & 200 \\
    MFIX & 120  & 120  & 120 & 120 & 120 \\
    \midrule
    \modelabb(C) & 36  & \underline{36}  & \underline{36} & 36 &  \underline{36} \\
    \modelabb(L) & \textbf{10}  & \textbf{10} & \textbf{10} & \textbf{10} & \textbf{10}  \\
    
    \bottomrule

    \end{tabular}
    \end{adjustbox}

\end{table}

\begin{table}[t]

    \caption{Average online recommendation time for different kinds of methods across 5 OLAP benchmarks (seconds).}
    \centering
    \vspace{-0.7em}
    \begin{adjustbox}{width=0.44\textwidth}
    \label{tab:runtime}

    \begin{tabular}{@{}lccccc@{}}
    
    \toprule
        & \textbf{TPC-H (SF = 5)} & \textbf{JOB} & \textbf{TPC-DS} & \textbf{SSB} & \textbf{TPC-H (SF = 50)} \\
    \midrule
    Heuristic methods & \textbf{10.9} & \underline{1897.5} & 1778.6 & \textbf{5.8} & \textbf{11.1} \\
    Learning-based methods & 139.9 & 7861.5 & 2961.3 & 109.2 & 186.3 \\
    \modelabb(C) & \underline{52.5}  & \textbf{89.2}  & \textbf{91.2}  & \underline{76.2} & \underline{84.6} \\
    \modelabb(L) & 1316.0  & 3555.0  & \underline{1687.5} & 1099.2 & 82,819.4 \\
    
    \bottomrule

    \end{tabular}
    \end{adjustbox}

\end{table}


\begin{table*}[t]
    \caption{Ablation studies on input features, demonstration number and LLM hyperparameters.}
    \centering
    \vspace{-0.8em}
    \begin{adjustbox}{width=0.8\textwidth}
    \label{tab:ablation_0}
    \begin{tabular}{@{}c|c|cccc|cc|c@{}}
    
    \toprule
    \multirow{2}{*}{Workload} & \multirow{2}{*}{\modelabb(L)} & \multicolumn{4}{c|}{Input Features} & \multicolumn{2}{c|}{Demonstrations Number} & LLM Hyperparameters \\
    \cmidrule(r){3-9}
    & & Raw Workload & Only SQL-Level Information & w/o WHERE Predicates & w/o JOIN Conditions & Zero-Shot & One-Shot & Temperature = 0\\
    \midrule
    TPC-H (SF = 5, 30\%) & 20.92 & 16.73 & 15.88 & 17.81 & 13.00 & 1.07 & 15.19 & 3.84 \\ 
 
    \bottomrule
        
    \end{tabular}

    \end{adjustbox}
\end{table*}

\subsection{Online Cost Evaluation}
\label{sec:eval_on_cost}
We assess online recommendation cost involving efficiency evaluation and LLM API's overhead evaluation (RQ2).

\subsubsection{Efficiency Analysis}
Given the different forms of external feedback used by existing methods and \model, we evaluate online recommendation efficiency by quantifying the number of benefit evaluations, including cost estimations and workload executions. We also report average online recommendation time across different methods, with the results summarized in Tables~\ref{tab:efficiency} and \ref{tab:runtime} (the best results highlighted in bold and the second-best underlined).
We have the following findings:

\textbf{\modelabb(L) exhibits superior efficiency with the minimum benefit evaluation requirements from the DBMS.} 
For heuristic methods, the benefit evaluation counts primarily depends on the number of index candidates and the search algorithm, with higher counts for sophisticated database schemas such as JOB and TPC-DS. For learning-based methods, the benefit evaluation counts are predefined by their hyperparameters, which can be calculated as: ${\text{Iteration Number per Epoch}} \times {\text{Training Epoch Number}}$. For our \model, the benefit evaluation counts can be calculated as: ${(\text{Multi-Sample Number} + 1)} \times {\text{Self-Refinement Number}}$. Experimental results illustrate our \model's minimal requirements of the external feedback, which could be attributed to the LLM's intrinsic reasoning capability that can directly determine the promising indexes without extensive iterative refinements. And this satisfactory efficiency enables \smodel directly utilizing workload actual latency for index recommendation, which exhibits satisfactory efficacy in Section~\ref{sec:main_results}.

\textbf{\modelabb(C) demonstrates an outstanding online recommendation time with the second minor database feedback.}
\modelabb(C) exhibits the superior and stable online recommendation time across different complexities of the workloads while maintaining almost second-best workload performance. Considering the limited accuracy of the cost estimation, we adaptively increase the number of obtaining external feedback via doubling the sample number (from 4 to 8) and iteration counts (from 2 to 4), which could relatively alleviate the misleading through expanding the candidate index sets.


\subsubsection{LLM Overhead Analysis}\label{sec:llm_overhead}
The LLM API overhead of \smodel can be calculated as: $\text{input tokens} \times \text{price per input token} + \text{output tokens} \times \text{price per output token}$. For \model, the input tokens are primarily dominated by the matched demonstrations and extracted workload features, while the output tokens are determined by the LLM’s generation sequence. GPT-4o-mini is priced at \$0.15 per 1 million input tokens and \$0.60 per 1 million output tokens. The detailed API cost of \smodel is calculated in Table~\ref{tab:overhead}.

\subsection{Ablation Study}
\label{sec:abla}
We conduct ablation experiments under latency for TPC-H SF = 5 with 30\% storage constraint to verify the effectiveness of \model(L)'s different components (RQ3).

\subsubsection{Ablations on Input Features}\label{ablation:input_features}
To assess the importance of workload features (detailed in Section~\ref{sec:wf}), we conducted fine-grained ablation studies by using raw workload or omitting specific condition features. 
As shown in Table~\ref{tab:ablation_0}, removing any of these features leads to a clear performance degradation, highlighting the critical role of comprehensive feature extraction in effective index recommendation.

\begin{table*}[t]
    \caption{Ablation studies on database feedback-driven selection and refinement and demonstration match strategies.}
    \centering
    \vspace{-0.7em}
    \begin{adjustbox}{width=0.8\textwidth}
    \label{tab:ablation_1}
    \begin{tabular}{@{}l|c|cccc|cc@{}}
    
    \toprule
    \multirow{2}{*}{Workload} & \multirow{2}{*}{\modelabb(L)} & \multicolumn{4}{c|}{Database Feedback-Guided Selection and Revision} & \multicolumn{2}{c}{Demonstration Match Strategies} \\
    \cmidrule(r){3-6}\cmidrule(r){7-8}
    & & Sample 1 + Ref 2 & Sample 2 + MV + Ref 2 & Sample 4 + Ref 2 & Sample 4 + MV + Ref 1 & Random Sample & K-Means Clustering \\
    \midrule
    TPC-H (SF = 5, 30\%) & 20.92 & 0.61 & 17.81 & 16.96 & 16.82 & 16.32 & 5.50 \\ 
    
    \bottomrule
        
    \end{tabular}

    \end{adjustbox}
\end{table*}

\begin{table}[t]
    \caption{Ablations on LLM backbones (TPC-H SF = 5 under 30\% storage constraint).}
    \centering
    \vspace{-0.7em}
    \begin{adjustbox}{width=0.48\textwidth}
    \label{tab:ablation_2}
    \begin{tabular}{@{}c|cccc@{}}
    
    \toprule
    \modelabb(L) & \multicolumn{4}{c}{LLM Backbones} \\
    \midrule
    & GPT-4.1-mini~\cite{gpt4.1} & DeepSeek-V3~\cite{deepseek} & LLaMA-3.3-70B~\cite{llama3_3} & LLaMA-3.1-8B~\cite{llama3_1} \\
    \cmidrule(r){2-5}
    \multirow{3}{*}{20.92} & 16.71 & 21.02 & 18.54 & 20.8 \\
    \cmidrule(r){2-5}
    & Qwen2.5-7B~\cite{qwen2_5} & Qwen2.5-Coder-7B~\cite{qwen2_5} & Qwen3-4B~\cite{qwen3} & Qwen3-30B-A3B~\cite{qwen3} \\
    \cmidrule(r){2-5}
    & 20.89 & 20.01 & 17.64 & 19.27 \\

    \bottomrule
        
    \end{tabular}

    \end{adjustbox}
\end{table}

\subsubsection{Ablations on Demonstrations}
\label{sec:abla_demos}
To tackle the lack of database expertise in LLMs, we use GPT-4-Turbo for workload generation and heuristic methods for demonstration annotation, as presented in Section~\ref{sec:demo}. We conduct ablation experiments from various perspectives to verify its effectiveness.

\vpara{Ablations on the Number of Demonstrations.}
We reduce the number of demonstrations to evaluate their necessity (see ``Demonstrations'' in Table~\ref{tab:ablation_0}). Experimental results show that performance degrades progressively as the number of demonstrations decreases, indicating LLMs are deficient in database expertise and highlighting the demonstrations' significance.

\vpara{Ablations on Workload Generation.}
Given that the workloads synthesized by LLMs lack professional quality assessment, we invite three expert DBAs to filter the demonstrations, retaining 50 as human-supervised demonstrations. For comparison, we also randomly select 50 demonstrations. We then evaluate \smodel using these two demonstration pool under TPC-H (SF = 5 with 30\% storage constraint).
The results show that human-supervised demonstrations slightly outperform the random set, achieving 17\% compared to 16.17\%, while neither reaches the performance of the original demonstration pool, which achieves 20.92\%. This indicates that increasing the quantity and diversity of demonstrations can reduce human supervision costs while still improving performance.

\vpara{Ablations on Refinement Demonstration Annotation.}
Due to the inaccuracy of executing \texttt{EXPLAIN} command for cost estimation, we re-annotate the refinement actions of these demonstrations using workload execution latency. Approximately 67.4\% of the refinement actions are updated. Experimental results under the updated demonstrations yield a performance of 20.12\%, slightly below the original 20.92\%, indicating that re-annotation does not lead to a significant improvement.
We speculate that could be owing to in-context learning fundamentally guides the LLM by enriching the external input information, rather than internalizing the content of the demonstrations into the model’s inherent capabilities. This also constitutes an advantage over supervised fine-tuning, which significantly reduces the cost of data synthesis.

\subsubsection{Ablations on LLM Hyperparameters}
We can control the diversity of the LLM's output content by adjusting its temperature, with a higher value denoting more diverse results with reduced reliability. 
We set the temperature to 0.8 in original \model, and evaluate the setting of the temperature = 0 to limit the LLM's exploration (see ``LLM Hyperparameters'' in Table~\ref{tab:ablation_0}). This leads to noticeable performance degradation, elucidating it infeasible to maintain the LLM's reliability through decreasing the hyperparameter of temperature.

\subsubsection{Ablations on Demonstrations Match Strategies}
\label{sec:abla4demomatch}
It is essential to select the most beneficial demonstrations, and we conduct an experiment comparing the different demonstration match strategies, as presented in Table~\ref{tab:ablation_1}. We observe the cosine similarity ranking strategy adopted in \smodel slightly outperforms other two strategies: random sampling and K-Means clustering, indicating the original method could present the most precise localization of the relevant demonstrations.




\subsubsection{Ablations on Database Feedback-Driven Selection and Refinement}
We introduce a database feedback-driven selection and refinement strategy (detailed in Section~\ref{sec:inferscaling}), which includes multi-sampling (abbreviated as Sample), major voting (abbreviated as MV), and a feedback-driven index refinement mechanism (abbreviated as Ref). Ablation studies presented in Table~\ref{tab:ablation_1} exhibit severe performance degradation when any components in this strategy are removed, underscoring the importance of the complete strategy.


\subsubsection{Ablations on LLM Backbones}
\label{sec:abla_llms}
To validate \model's robustness, we replaced GPT-4o-mini with various LLM models, including the GPT-4.1-mini~\cite{gpt4technical}, DeepSeek-V3~\cite{deepseek}, LLaMA-3.3-70B~\cite{llama3_3}, LLaMA-3.1-8B~\cite{llama3_1}, Qwen2.5-7B~\cite{qwen2_5}, Qwen2.5-Coder-7B~\cite{qwen2_5}, Qwen3-4B~\cite{qwen3}, and Qwen3-30B-A3B~\cite{qwen3}. Experimental results presented in Table~\ref{tab:ablation_2} demonstrate that \model's well-designed framework can effectively adapt to different LLM models for promising index recommendation, highlighting our framework's versatility.

\subsubsection{Ablations on \model's stability}
\label{sec:abla4stab}
Considering the uncertainty of LLM inference, we perform index recommendations under TPC-H (SF = 5, 30\% storage constraint) across 10 trials to assess LLM inference's variability.
The results show that \smodel achieves maximum and average latency reductions of 22.14\% and 18.23\%, respectively, outperforming all baselines. Even the minimum reduction of 16.55\% exceeds most baselines, indicating \model's reliable index recommendations.

\subsection{Analysis of Offline Preparation}
\label{eval_off_cost}

We additionally analyze the offline preparation (RQ4). Heuristic methods directly produce solutions via greedy search over index candidates, incurring little offline overhead but substantial online recommendation time. Learning-based methods typically require numerous iterations before applying to a workload, which could be treated as an online recommendation. Both approaches may fail to deliver satisfactory workload performance due to inaccurate cost estimation.
For \model, we briefly outline the cost involved in the offline demonstration construction stage. 
For workload generation, invoking the GPT-4-Turbo API incurs costs of \$10 per million tokens for input and \$30 per million tokens for output. For the TPC-H, JOB, and TPC-DS database schemas, the total input token consumptions are approximately 1,454,250, 1,196,838, and 2,270,065 tokens, respectively, while the corresponding output token consumptions total 396,874, 413,730, and 552,345 tokens. The cumulative cost amounts to \$49.21 for input tokens and \$40.89 for output tokens. Using GPT-4-Turbo, we sample about 1,000 SQL queries per database, which takes nearly 17 hours for 3 database schemas.
For label annotation, we employ multiple heuristic methods to generate candidate actions. This process takes about 2 min per workload on TPC-H, while requiring 1 hour for JOB and TPC-DS, finally distributing on 10 CPU servers spending within a single day for all synthetic workloads.
Notably, this offline preparation is a one-time process, which could be directly applied out-of-the-box to diverse scenarios without repeating this effort. 
Experimental results in Figure~\ref{fig:rw} demonstrate that \modelabb(L) with minimal database feedback exhibits superior efficacy for unseen workloads. This also suggests that the primary role of demonstrations is to impart database expertise to the LLM, while the core information for LLM-based index recommendation comes from the extracted feature of the target workload and accurate feedback obtained from the DBMS.

\section{Discussion}
\label{sec:dis}
We conducted an additional pilot study to explore whether supervised fine-tuning (SFT) of LLMs under equivalent data requirement is a superior alterative for index recommendation.
\begin{table}[t]

    \caption{Latency evaluation of fine-tuning under 30\% storage constraint. The symbol ``-'' means evaluation timeout.
    } 
    \centering
    \vspace{-0.6em}
    \begin{adjustbox}{width=0.48\textwidth}
    \label{tab:sft}

    {
    \begin{tabular}{@{}lccccc@{}}

    \toprule
        & \textbf{TPC-H (SF = 5)} & \textbf{JOB} & \textbf{TPC-DS} & \textbf{SSB} & \textbf{TPC-H (SF = 50)} \\
    \midrule
    \modelabb(L) & \textbf{20.92} & \textbf{76.46} & \underline{15.52} & \underline{13.93} & \textbf{7.97} \\
    \midrule
    \modelabb(L) w/ LLaMA-3.1-8B & \underline{20.8} & \underline{75.59} & \textbf{17.78} & 13.57 & \underline{6.64} \\
    \midrule
    Qwen2.5-7B & 5.42 & 20.49 & 1.08 & 11.83 & 2.87  \\
    Qwen2.5-Coder-7B & 0.64 & 31.53 & 12.41 & 4.89 & - \\
    LLaMA-3.1-8B & 5.53 & 72.48 & -5.37 & \textbf{15.81} & 1.17 \\
    \bottomrule

    \end{tabular}
    }
    
    \end{adjustbox}

\end{table}


\subsection{Environment and Data Preparation}
We select Qwen2.5-7B-Instruct~\cite{qwen2_5}, Qwen2.5-Coder-7B-Instruct~\cite{qwen2_5}, and Meta-Llama-3.1-8B-Instruct~\cite{llama3_1} for fully supervised fine-tuning, with the learning rate of 1e-5, the batch size of 24, and 8 epochs. 
Both model training and inference are conducted on a single machine equipped with 3 NVIDIA H20-SXM5-96GB GPUs, and the training process takes approximately 4 hours per model.
For fairness, we take 590 original demonstrations for training. Since each demonstration contains only workload features and refinement action—without iterative database-feedback signals (e.g., historical information)—they cannot be directly standardized into \model's reasoning format. We simply modify training data pairs where the input is the system instruction plus workload features, and the output is the initial refinement action.


\subsection{Performance Evaluation}
We recommend indexes using a fine-tuned LLM with the temperature to 0 and the sample number to 1, and conduct a comparable experiment using LLaMA-3.1-8B with \model’s framework. Experimental results are presented in Table~\ref{tab:sft}.
Although fine-tuned model consistently achieves faster recommendation times (within one minute), \model’s tuning-free framework with different LLM backbones still delivers superior performance, which can be attributed to two key factors.
First, the demonstrations’ initial refinement actions are annotated using estimated costs, which can differ markedly from execution latencies (refer to Section~\ref{sec:abla_demos}). Therefore, supervised fine-tuning may overfit to these limited and potentially inaccurate annotations. Second, the fine-tuned model recommends indexes via a single inference, lacking the ability to incorporate database feedback for iterative refinement.
In contrast, \smodel simulates the DBAs' decision-making process by leveraging its database feedback–driven, iterative recommendation framework to identify promising indexes.

\subsection{Discussion}
\label{sec:dis_dis}
\smodel is an LLM-based tuning-free index advisor equipped with a pre-constructed demonstration pool. In this paper, we primarily target the read-heavy scenarios. \smodel possesses the LLM's interpretability, which can be regarded as a copilot to assist DBAs in managing the DBMS. 

\subsubsection{Application}
\label{sec:dis_app}
Experimental results illustrate \modelabb(L) achieves superior performance with lengthy recommendation time, while \modelabb(C) maintains higher efficiency with appropriate performance. Specifically, 
\modelabb(L) is designed for workloads with recurring workloads with fixed indexes demands (e.g., weekly executed financial reporting queries, analytical workloads with pre-defined aggregation patterns, etc.).
\modelabb(C) could be suitable for workloads with real-time requirements or urgent demands (e.g., emergent fraud detection queries, interactive customer-facing search requests, etc.). 
Since \modelabb(L) requires actual index modifications, a practical alternative is to apply it on a temporary database replica for index recommendation, improving query latency without disrupting the dynamic workload scenarios.

\subsubsection{Implementation}
\label{sec:dis_imp}
\smodel can be treated as an external tool that recommends indexes without requiring integration with the DBMS and supports different LLM backbones for inference. When the input information exceeds the LLM’s context length, \smodel maintains maximal recommendation performance by sequentially trimming SQL-level details, demonstrations, and workload-level features.
Although this paper focuses on index recommendation for read-heavy workloads in PostgreSQL, \smodel can also be adapted for write-heavy scenarios. For example, a stress-testing tool can be used for write-heavy workload execution while the actual index maintenance cost will be considered in this process, 
and database statistics can be refreshed according to data update schedules to improve input information's accuracy.
In summary, \modelabb’s well-designed index recommendation framework can be applied to various database systems through replacing the corresponding components according to the specific system's characteristics.
\section{Conclusion}
This paper presents \model, an out-of-the-box, tuning-free index advisor powered by large language models (LLMs) and in-context learning. \smodel emulates DBAs' decision-making process, leveraging a curated demonstration pool, comprehensive workload feature extraction, and iterative database feedback to refine indexes efficiently and accurately. 
Extensive experiments on standard OLAP benchmarks and real-world benchmarks demonstrate \smodel achieves superior or comparable performance to strong baselines, with minimal DBMS interaction and robust out-of-the-box generalization capability.

\section*{Acknowledgment}
This work is supported by the National Key Research \& Development Plan (2023YFF0725100) and the National Natural Science Foundation of China (92570121, 62322214, U23A20299, U24B20144). We also acknowledge the support of the Public Computing Cloud, Renmin University of China.

\section*{AI-Generated Content Acknowledgement}

Portions of the English writing in this article were refined with the assistance of OpenAI’s ChatGPT, a large language model. ChatGPT was used to help improve the clarity and fluency of some sentences across the Introduction, Related Work, and Method sections. All technical content, experimental design, analysis, and scientific conclusions were generated by the authors.

\bibliographystyle{plain}
\bibliography{_reference}

\end{document}